\documentclass[journal]{IEEEtran}
\ifCLASSINFOpdf
\else
\fi

\usepackage{graphicx}
\usepackage{epstopdf}
\usepackage{epsfig}

\usepackage{amsmath,amsfonts}
\usepackage{latexsym, bm}

\usepackage{mathrsfs}
\usepackage{amsbsy}

\usepackage{graphicx}
\usepackage[caption=false]{subfig}

\usepackage{algorithm}
\usepackage{algpseudocode}

\usepackage{cite}
\usepackage{stmaryrd}

\usepackage{tabularx}

\begin{document}

\newcolumntype{L}[1]{>{\raggedright\arraybackslash}p{#1}}
\newcolumntype{C}[1]{>{\centering\arraybackslash}p{#1}}
\newcolumntype{R}[1]{>{\raggedleft\arraybackslash}p{#1}}
%
\title{Two Types of Mixed Orthogonal Frequency Division Multiplexing (X-OFDM) Waveform for Optical Wireless Communication}
%
%
%

\author{Xu Li, Jingjing Huang, Yibo Lyu, Rui Ni, Jiajin Luo, Junping Zhang 
\thanks{X. Li, J. Huang, Y. Lyu, R. Ni, J. Luo, J. Zhang are with the Shenzhen Research Development Center, Huawei Technologies Co., Ltd., 518129, China.}
\thanks{Corresponding author: X. Li, e-mail: lixu11@huawei.com.}
}

%
%

\markboth{Journal of \LaTeX\ Class Files,~Vol.~XX, No.~XX, October~2020}%
{Shell \MakeLowercase{\textit{et al.}}: Bare Demo of IEEEtran.cls for IEEE Journals}
%



\maketitle

\begin{abstract}
The optical wireless communication (OWC)  with the intensity modulation (IM), requires the modulated radio frequency (RF) signal to be real and non-negative. To satisfy the requirements, this paper proposes two types of  mixed orthogonal frequency division multiplexing (X-OFDM) waveform. The Hermitian symmetry (HS) character of the sub-carriers in the frequency domain, guarantees the signal in the time domain to be real, which reduces the spectral efficiency to $1/2$. For the odd sub-carriers in the frequency domain, the signal in the time domain after the inverse fast fourier transform (IFFT) is antisymmetric. For the even sub-carriers in the frequency domain, the signal in the time domain after the IFFT is symmetric. Based on the antisymmetric and symmetric characters, the two types of  X-OFDM waveform are designed to guarantee the signal in the time domain to be non-negative, where the  direct current (DC) bias is not needed. With $N$ sub-carriers in the frequency domain, the generated  signal in the time domain has $3N/2$ points, which further reduces the spectral efficiency to $1/3 = 1/2 \times 2/3$. The numerical simulations show that, the two types of X-OFDM waveform greatly enhance the power efficiency considering the OWC channel with the signal-dependent noise and/or the signal-independent noise.
\end{abstract}

\begin{IEEEkeywords}
Optical wireless communication (OWC), Intensity Modulation (IM), Mixed Orthogonal Frequency Division Multiplexing (X-OFDM) Waveform, Peak to Average Power Ratio (PAPR), Bit Error Rate (BER), Power Efficiency
\end{IEEEkeywords}

%
\IEEEpeerreviewmaketitle

\begin{figure*}[]
\centering
\includegraphics[width=0.99\textwidth]{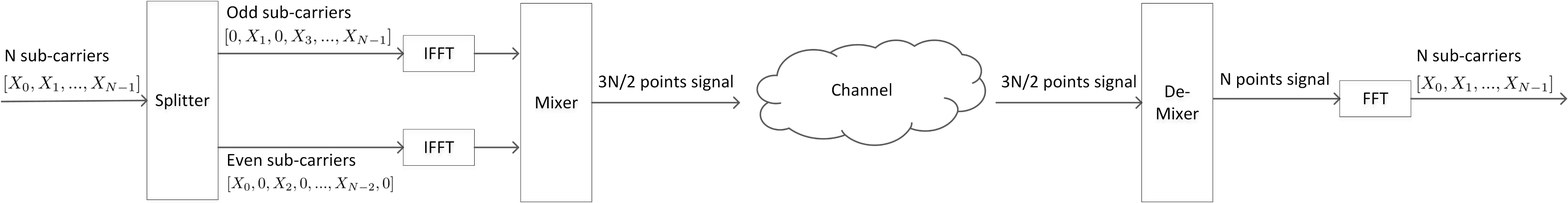}
\caption{The system architecture of the first type of X-OFDM waveform.}
\label{SYSTEMMODEL}
\end{figure*}

\section{Introduction}
The optical wireless communication (OWC) is a candidate technology for the communication beyond fifth-generation (5G),  to provide the services with ultra-high system capacity, ultra-low latency, ultra-high security, ultra-low energy consumption, and etc. \cite{6G}. Differs from the wireless radio frequency (RF) band between 3kHz and 300GHz of the electromagnetic spectrum, the wireless connectivity based on the optical spectrum is termed as the OWC \cite{OWC}. 

For the optical communication, there exist the coherent and the non-coherent communication schemes. With the coherent scheme, both the intensity and the phase of the optical carrier carry information. However, the system architecture is complicated and the related optical devices are expensive. The scheme is usually used in the optical fiber communication system \cite{coherent}. With the non-coherent scheme, only the intensity of the optical carrier carries information, which is  named as the intensity modulation (IM) and adopted in the OWC system \cite{OWC,VLC}. The intensity of the optical carrier is real and non-negative, hence also the modulated RF signal.


The orthogonal frequency division multiplexing (OFDM) waveform is a key feature in the 4G and the 5G RF communication \cite{slicing}, which is also utilized in IEEE 802.15.13, IEEE 802.11b and ITU-9991 standards for the OWC \cite{IEEE13,IEEE11,ITU}. To satisfy the real and non-negative requirements, the OFDM waveform is modified as the DC-biased optical OFDM (DCO-OFDM) waveform, the unipolar OFDM (U-OFDM) waveform and the asymmetrically clipped optical OFDM (ACO-OFDM) waveform. The Hermitian symmetry (HS) character of the signal in the frequency domain, guarantees the signal in the time domain to be real at the sacrifice of 1/2 spectral efficiency. To be non-negative, the DCO-OFDM waveform adds the DC bias and clips the negative parts with 1/2 spectral efficiency. Without the DC bias, the U-OFDM waveform flips the negative parts following the non-negative parts, which doubles the signal length in the time domain with 1/4 spectral efficiency.  The ACO-OFDM waveform only uses the odd sub-carriers in the frequency domain with 1/4 spectral efficiency. Then the signal in the time domain is antisymmetric, and the negative parts are clipped directly \cite{book}.

To further increase the spectral efficiency and reduce the power consumption, the hybrid approach as the asymmetrically clipped DC biased optical OFDM (ADO-OFDM) waveform \cite{ADO-OFDM} and the superposition approach as the layered ACO-OFDM waveform \cite{L-ACO-OFDM} and the enhanced U-OFDM waveform\cite{E-U-OFDM} are analyzed. The ADO-OFDM waveform combines the DCO-OFDM at the even sub-carriers and the ACO-OFDM at the odd sub-carriers. The layered ACO-OFDM waveform has multiple layers, where each layer uses the carefully selected sub-carriers. The enhanced U-OFDM waveform has multiple depths, where each depth repeats the signal multiple times. However, all the receivers require the successive demodulation scheme, where part of the signal is demodulated and reconstructed to erase the interference, before other parts of the signal to be demodulated. The receiver is complicated and suffers from the error propagation problem \cite{E-P}.

In this paper, two types of mixed OFDM (X-OFDM) waveform without the DC bias are proposed, where the first type of X-OFDM waveform uses the non-successive demodulation scheme. Similarly, for the signal in the time domain to be real, the $N$ sub-carriers in the frequency domain satisfy the HS requirement. Then the sub-carriers are divided into the odd sub-carriers and the even sub-carriers. For the odd sub-carriers, after the inverse fast fourier transform (IFFT), the signal in time domain is antisymmetric, where the first $N/2$ points are the inverse of the second $N/2$ points. For the even sub-carriers, after the IFFT, the signal in the time domain is symmetric, where the first $N/2$ points equal to the second $N/2$ points. Then there are four independent $N/2$ points extracted as: 1) the non-negative parts of the first $N/2$ points generated from the odd sub-carriers; 2) the negative parts of the first $N/2$ points generated from the odd sub-carriers; 3) the non-negative parts of the first $N/2$ points generated from the even sub-carriers; 4) the negative parts of the first $N/2$ points generated from the even sub-carriers. The four independent $N/2$ points are mixed to be a $3N/2$ points sequence at the transmitter and de-mixed to be decoded at the receiver in different ways for the two types of X-OFDM waveform.   The spectral efficiency, computational complexity, motivation and comparison are discussed, and the numerical simulations are provided to evaluate the performance of the two types of X-OFDM waveform.

The first type of X-OFDM waveform is briefly introduced in our conference paper \cite{globalcom}, but comprehensively studied along with the second type of X-OFDM waveform in this paper. The contributions of this paper are summarized as:
\begin{itemize}
\item Two types of X-OFDM waveform without the DC bias are proposed for the OWC with the IM, where the modulated RF signal is real and non-negative;
\item The related system architecture, signal processing, spectrum efficiency, computational complexity, motivation and comparison, and numerical simulations considering the relative intensity noise, shot noise, background radiation noise and thermal noise of the OWC channel \cite{channel1} are comprehensively studied;
\item The first type of X-OFDM waveform uses the non-successive demodulation scheme, which has $1/3$ spectral efficiency, $\mathcal{O}(3N\log_2N)$ computational complexity, low peak to average power ratio (PAPR) and high power efficiency especially for the scenarios with the high relative intensity noise and shot noise;
\item The second type of X-OFDM waveform uses the successive demodulation scheme, which has $1/3$ spectral efficiency, $\mathcal{O}(5N\log_2N)$ computational complexity, higher PAPR but higher power efficiency than the first type of X-OFDM waveform especially for  the scenarios with the high relative intensity noise and shot noise.
\end{itemize}


The remainder of this paper is organized as follows. Section \ref{section2} introduces the first type of X-OFDM waveform with the system architecture and the signal processing at the transmitter and the receiver. Section \ref{section22} introduces the second type of X-OFDM waveform with the system architecture and the signal processing at the transmitter and the receiver. The spectral efficiency, computational complexity, motivation and comparison are discussed in Section \ref{section3}.  Section \ref{section4} presents the numerical simulations, followed by the conclusions in Section \ref{section5}.

\begin{figure}[]
\centering
\subfloat[]{
\label{fig:p1}\includegraphics[width=0.45\textwidth]{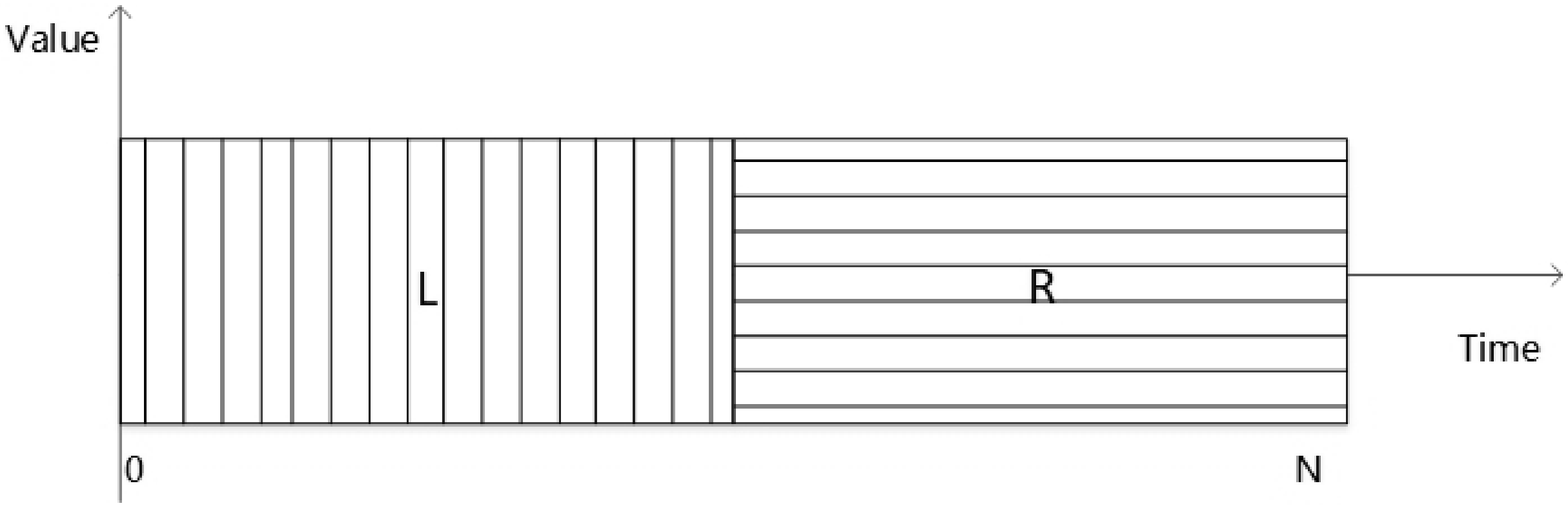}} \\
\subfloat[]{
\label{fig:p2}\includegraphics[width=0.45\textwidth]{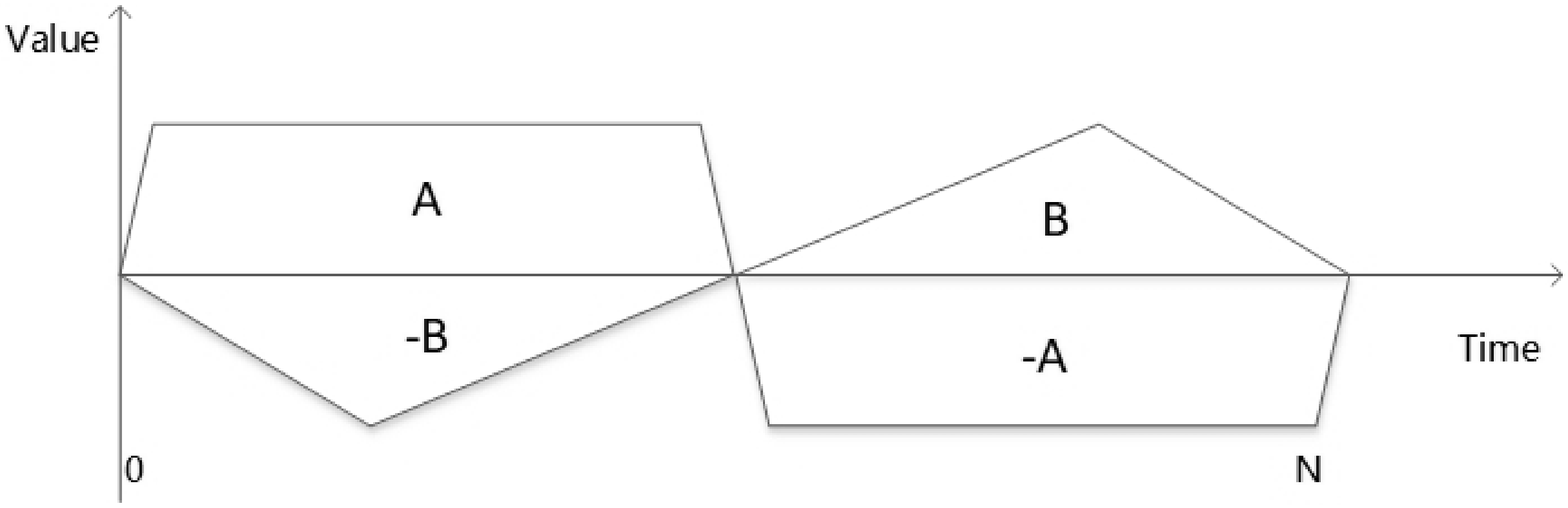}} \\
\subfloat[]{
\label{fig:p3}\includegraphics[width=0.45\textwidth]{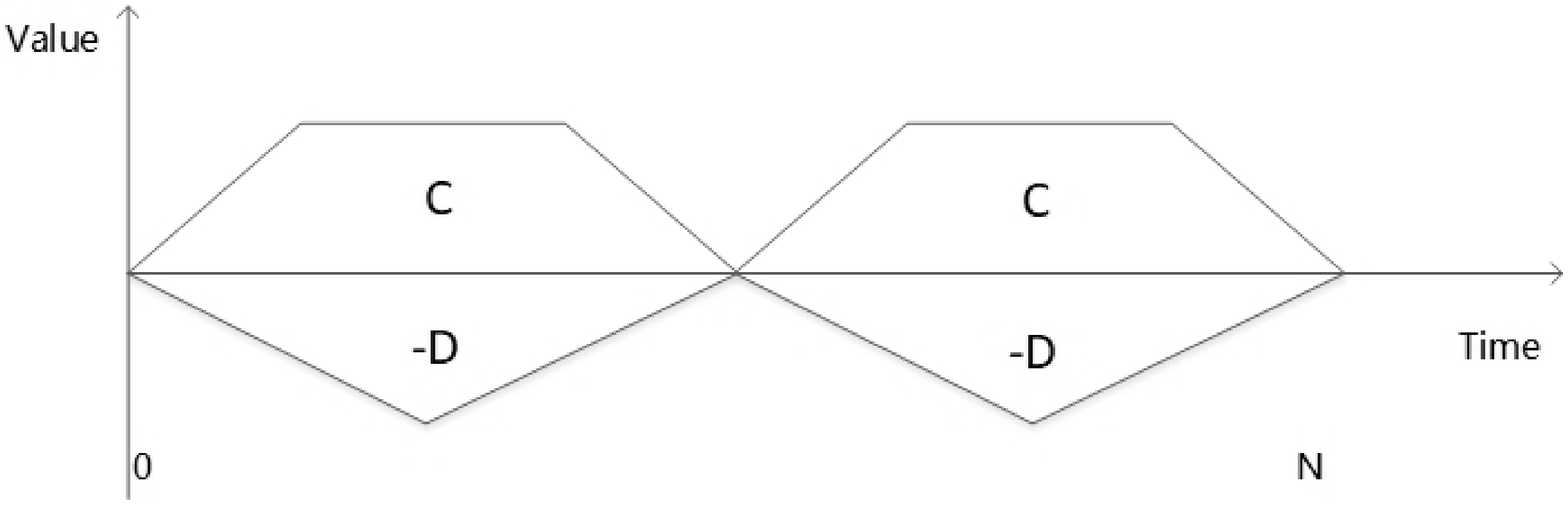}} \\
\subfloat[]{
\label{fig:p4}\includegraphics[width=0.38\textwidth]{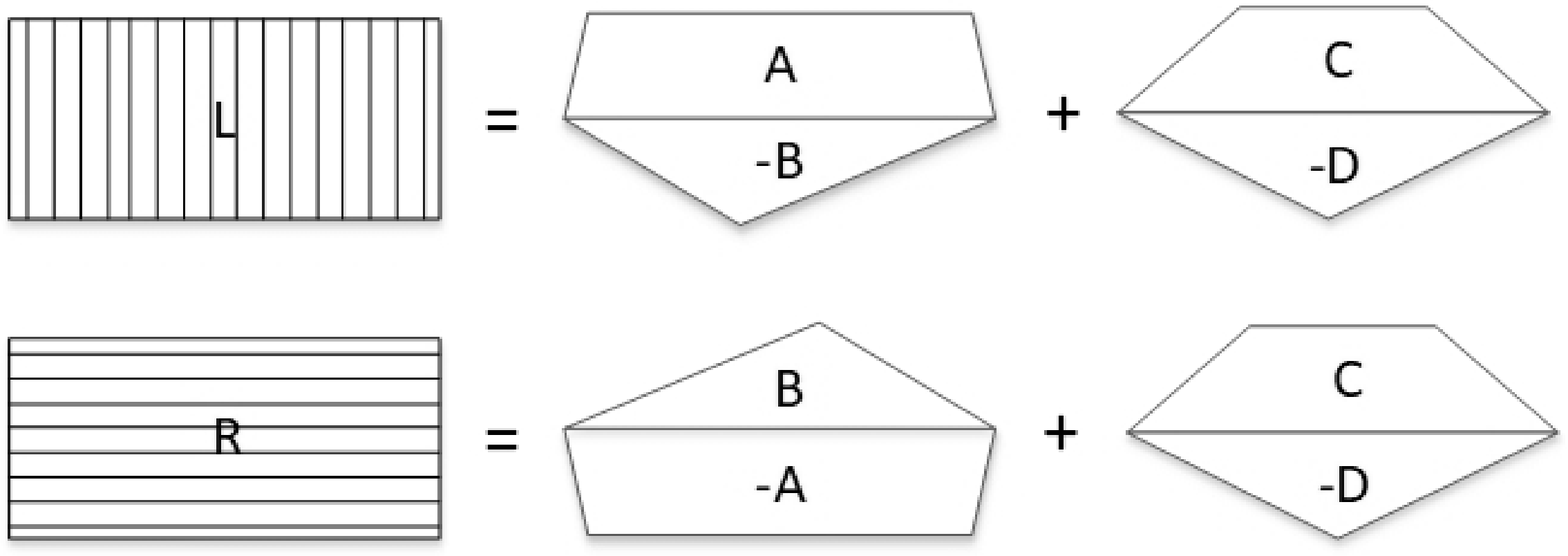}} \\
\subfloat[]{
\label{fig:p5}\includegraphics[width=0.43\textwidth]{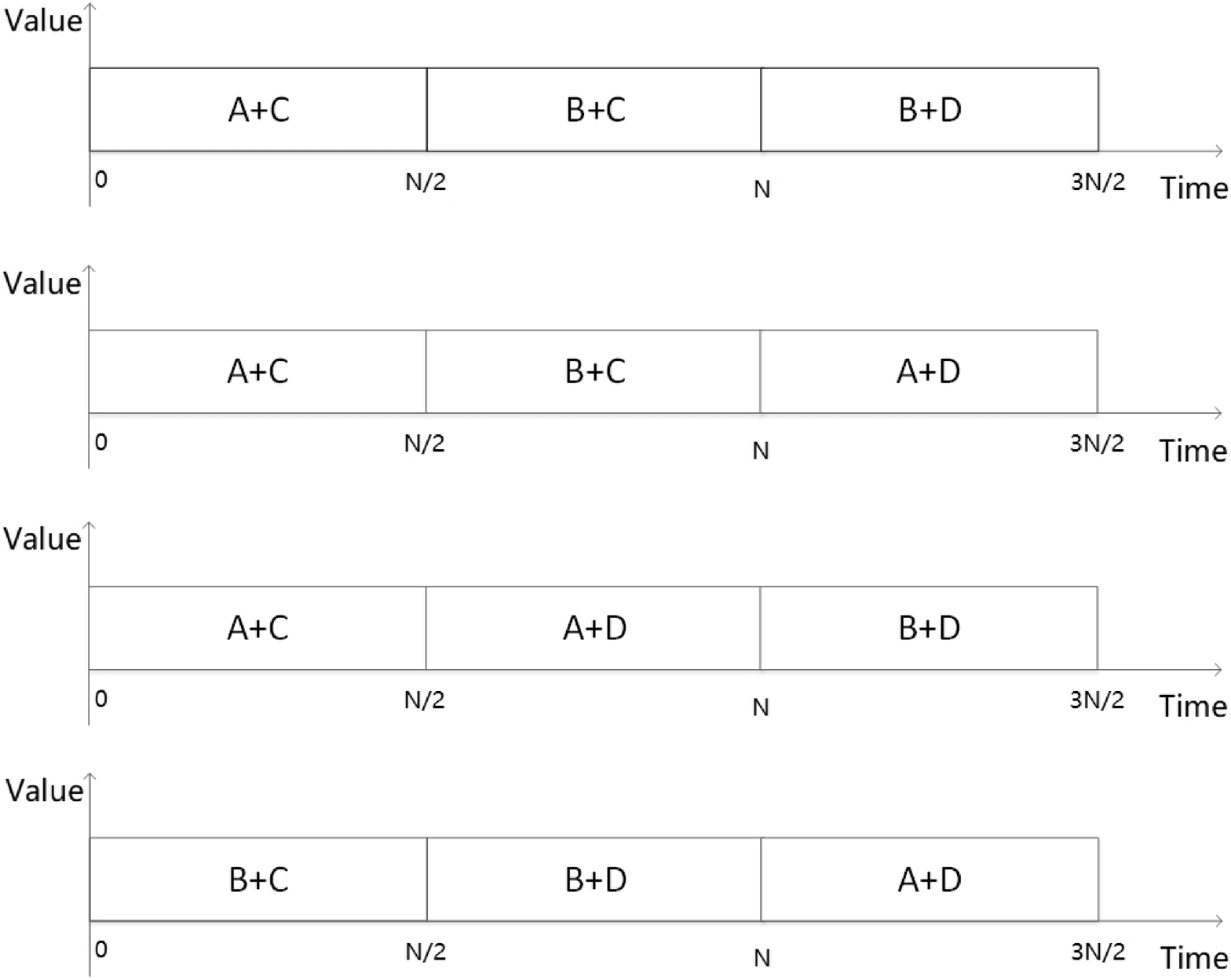}}
\caption{The signal processing at the transmitter of the first type of the X-OFDM waveform: (a) the real signal in the time domain with the sub-carriers satisfying the HS requirement, b) the antisymmetric signal in the time domain with the odd sub-carriers, (c) the symmetric signal in the time domain with the even sub-carriers, (d) the relationship of the signal in the time domain, (e) the candidate transmitted signal in the time domain.}
\label{fig:transmitter}
\end{figure}

\section{The First Type of X-OFDM Waveform}\label{section2}
The first type of X-OFDM waveform guarantees the signal in the time domain to be real and non-negative, without the DC bias and with the non-successive demodulation scheme. This section studies the system architecture and the signal processing at the transmitter and the receiver of the first type of X-OFDM waveform.

\subsection{System Architecture}
The system architecture of the first type of X-OFDM waveform is shown in Fig. \ref{SYSTEMMODEL}. At the transmitter, there exist a splitter, two IFFT blocks, and a mixer. The splitter splits the $N$ sub-carriers into the odd sub-carriers and the even sub-carriers. The IFFT blocks transform the signal into the time domain, followed by the mixer to generate a $3N/2$ points signal. At the receiver, there exist a de-mixer and a FFT block. Based on the $3N/2$ points signal, the de-mixer constructs a $N$ points signal, followed by the FFT block to recover the $N$ sub-carriers.

\subsection{Signal Processing at Transmitter}
The OWC  with the IM, requires the modulated RF signal to be real and non-negative. To be real, the HS character of the sub-carriers in the frequency domain $[X_0, X_1, ..., X_{N-1}]$ is used as \cite{book}:
\begin{align}
&X_m = X^*_{N-m}, 0<m<\frac{N}{2}.\notag \\
&X_0 = X_{\frac{N}{2}} = 0 \label{HS_equal}
\end{align}
and the signal in the time domain is:
\begin{align}
x_k^{origin} &= \sum_{m=0}^N X_m \exp\left(j\frac{2\pi}{N} mk\right) \notag \\
&= \text{Real} \left\{ 2\sum_{m=0}^{\frac{N}{2}} X_m \exp\left(j\frac{2\pi}{N} mk\right) \right\} \label{IFFT_NO}
\end{align}
which is shown in Fig.\ref{fig:p1}.  The first $N/2$ points are defined as $L$, and the second $N/2$ points are defined as $R$, where both $L$ and $R$ are real. We note that the IFFT equation (\ref{IFFT_NO}) and Fig.\ref{fig:p1} are shown for easier understanding of the X-OFDM waveform, but not needed in the real system.

The splitter splits the $N$ sub-carriers into the odd sub-carriers as $[0, X_1, 0, X_3, ..., X_{N-1}]$ and the even sub-carriers as $[X_0, 0, X_2, 0, ..., X_{N-2}, 0]$. For the odd sub-carriers, after the IFFT block, the signal in the time domain satisfies:
\begin{align}
&x_{k+\frac{N}{2}}^{odd} = \sum_{m=0}^{\frac{N}{2}-1} X_{2m+1} \exp\left(j\frac{2\pi}{N} (2m+1)(k+\frac{N}{2})\right)\notag\\
&= \sum_{m=0}^{\frac{N}{2}-1} X_{2m+1} \exp\left(j\frac{2\pi}{N} (2m+1)k\right) \exp \left(j \pi (2m+1)\right) \notag\\
&= -x_k^{odd}\label{AB_IFFT}
\end{align}
which is antisymmetric as shown in Fig.\ref{fig:p2}. For the first $N/2$ points, the non-negative parts are defined as $A$, and the negative parts are defined as $-B$, where both $A = [x_0^A, x_1^A, ..., x_{\frac{N}{2}-1}^A ]$ and $B = [x_0^B, x_1^B, ..., x_{\frac{N}{2}-1}^B ]$ are real and non-negative as:
\begin{align}
&x_{k}^{A} = \frac{|x_k^{odd}| + x_k^{odd}}{2},  0\leq k < \frac{N}{2}\label{A}\\
&x_{k}^{B} = \frac{|x_k^{odd}| - x_k^{odd}}{2},  0\leq k < \frac{N}{2}\label{B}
\end{align}

For the even sub-carriers, after the IFFT block, the signal in the time domain satisfies:
\begin{align}
&x_{k+\frac{N}{2}}^{even} = \sum_{m=0}^{\frac{N}{2}-1} X_{2m} \exp\left(j\frac{2\pi}{N} (2m)(k+\frac{N}{2})\right)\notag\\
&= \sum_{m=0}^{\frac{N}{2}-1} X_{2m+1} \exp\left(j\frac{2\pi}{N} (2m)k\right) \exp \left(j \pi (2m)\right) \notag\\
&= x_k^{even}
\end{align}
which is symmetric as shown in Fig.\ref{fig:p3}. For the first $N/2$ points, the non-negative parts  are defined as $C$, and the negative parts are defined as $-D$, where both $C = [x_0^C, x_1^C, ..., x_{\frac{N}{2}-1}^C ]$ and $D = [x_0^D, x_1^D, ..., x_{\frac{N}{2}-1}^D ]$ are real and non-negative as:
\begin{align}
&x_{k}^{C} = \frac{|x_k^{even}| + x_k^{even}}{2},  0\leq k < \frac{N}{2}\label{C}\\
&x_{k}^{D} = \frac{|x_k^{even}| - x_k^{even}}{2},  0\leq k < \frac{N}{2}\label{D}
\end{align}

Then there exist the relationships as shown in Fig.\ref{fig:p4} that:
\begin{align}
&L = (A+C) - (B+D)\notag\\
&R = (B+C) - (A+D) \label{LR}
\end{align}
and
\begin{align}
(A+C) + (B+D) = (B+C) + (A+D). \label{ABCD}
\end{align}
Based on the relationships, we could transmit either three of $(A+C)$, $(B+D)$, $(B+C)$, $(A+D)$ as shown in Fig.\ref{fig:p5}, and recover the remaining one using equation (\ref{ABCD}). The function of the mixer block is summarized as:
\begin{itemize}
\item The $A$, $B$, $C$, $D$ are extracted using the equations (\ref{A}), (\ref{B}), (\ref{C}) and (\ref{D});
\item The $A$, $B$, $C$, $D$ are mixed to construct either three of $(A+C)$, $(B+D)$, $(B+C)$, $(A+D)$ as shown in Fig.\ref{fig:p5}.
\end{itemize}
Then the constructed $3N/2$ points are transmitted at the transmitter.
\begin{figure*}[]
\centering
\includegraphics[width=0.99\textwidth]{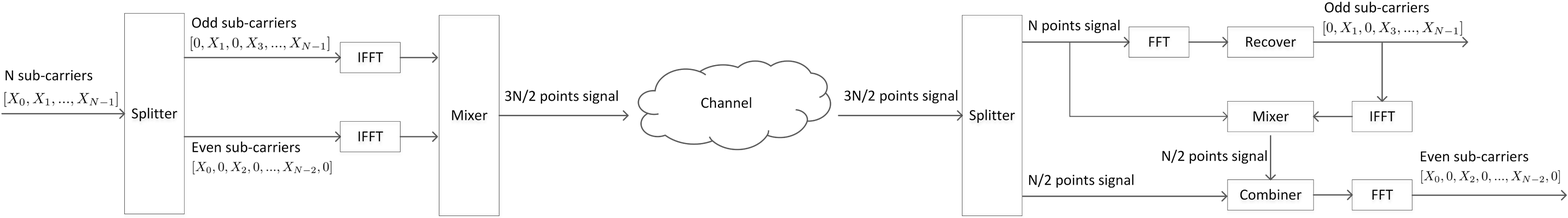}
\caption{The system architecture of the second type of X-OFDM waveform.}
\label{SYSTEMMODEL2}
\end{figure*}

\subsection{Signal Processing at Receiver}
The receiver receives the $3N/2$ points of either three of $(A+C)$, $(B+D)$, $(B+C)$, $(A+D)$ as shown in Fig.\ref{fig:p5}. Based on the received signal, the de-mixer constructs the $N$ points signal as shown in Fig.\ref{fig:p1} as:
\begin{itemize}
\item The remaining one of $(A+C)$, $(B+D)$, $(B+C)$, $(A+D)$ is recovered using equation (\ref{ABCD});
\item The $L$ and $R$ are constructed using equation (\ref{LR}).
\end{itemize}

The $N$ points signal as shown in Fig.\ref{fig:p1} goes through the FFT block to recover the $N$ sub-carriers in the frequency domain as $[X_0, X_1, ..., X_{N-1}]$.

\subsection{An Example}
A simple example is provided for the first type of X-OFDM waveform.
At the transmitter, the sub-carrier number is $N = 8$ and the constellation is 16-QAM. The signal of the sub-carriers are:
\begin{table}[!h]
\centering
\label{parameters}
\renewcommand{\arraystretch}{1.0}
\begin{tabular}{C{0.447\textwidth}}
\hline
$N=8$ sub-carriers $[X_0, X_1, ..., X_{7}]$ satisfying the HS requirement
\end{tabular}
\begin{tabular}{|C{0.035\textwidth}|C{0.035\textwidth}|C{0.035\textwidth}|C{0.035\textwidth}|C{0.035\textwidth}|C{0.035\textwidth}|C{0.035\textwidth}|C{0.035\textwidth}|}
\hline
0 &-3-j & -3+j & -1+3j & 0 & -1-3j & -3-j & -3+j\\
\hline
\end{tabular}
\end{table}
\\
The odd sub-carriers and $A$, $B$ are:
\begin{table}[!h]
\centering
\label{parameters}
\renewcommand{\arraystretch}{1.0}
\begin{tabular}{C{0.447\textwidth}}
\hline
The odd sub-carriers $[0, X_1, 0, X_3, ..., 0, X_{7}]$
\end{tabular}
\begin{tabular}{|C{0.035\textwidth}|C{0.035\textwidth}|C{0.035\textwidth}|C{0.035\textwidth}|C{0.035\textwidth}|C{0.035\textwidth}|C{0.035\textwidth}|C{0.035\textwidth}|}
\hline
0 &-3-j & 0 & -1+3j & 0 & -1-3j & 0 & -3+j\\
\hline
\end{tabular}
\begin{tabular}{C{0.447\textwidth}}
After the IFFT block to get $[x_0^{odd}, x_1^{odd}, ..., x_7^{odd}]$
\end{tabular}
\begin{tabular}{|C{0.035\textwidth}|C{0.035\textwidth}|C{0.035\textwidth}|C{0.035\textwidth}|C{0.035\textwidth}|C{0.035\textwidth}|C{0.035\textwidth}|C{0.035\textwidth}|}
\hline
-1 &-0.71& 1 & 0 & 1 & 0.71 & -1 & 0\\
\hline
\end{tabular}
\begin{tabular}{C{0.21\textwidth}|| C{0.21\textwidth}}
A & B
\end{tabular}
\begin{tabular}{|C{0.034\textwidth}|C{0.034\textwidth}|C{0.034\textwidth}|C{0.034\textwidth}||C{0.034\textwidth}|C{0.034\textwidth}|C{0.034\textwidth}|C{0.034\textwidth}|}
\hline
0 &0& 1 & 0 & 1 & 0.71 & 0 & 0\\
\hline
\end{tabular}
\end{table}
\\
The even sub-carriers and $C$, $D$ are:
\begin{table}[!h]
\centering
\label{parameters}
\renewcommand{\arraystretch}{1.0}
\begin{tabular}{C{0.447\textwidth}}
\hline
The even sub-carriers $[X_0, 0, X_2, 0, ..., X_6, 0]$
\end{tabular}
\begin{tabular}{|C{0.035\textwidth}|C{0.035\textwidth}|C{0.035\textwidth}|C{0.035\textwidth}|C{0.035\textwidth}|C{0.035\textwidth}|C{0.035\textwidth}|C{0.035\textwidth}|}
\hline
0 &0 & -3+j & 0 & 0 & 0 & -3-j & 0\\
\hline
\end{tabular}
\begin{tabular}{C{0.447\textwidth}}
After the IFFT block to get $[x_0^{even}, x_1^{even}, ..., x_7^{even}]$
\end{tabular}
\begin{tabular}{|C{0.035\textwidth}|C{0.035\textwidth}|C{0.035\textwidth}|C{0.035\textwidth}|C{0.035\textwidth}|C{0.035\textwidth}|C{0.035\textwidth}|C{0.035\textwidth}|}
\hline
-0.75 &-0.25& 0.75 & 0.25 & -0.75 & -0.25 & 0.75 & 0.25\\
\hline
\end{tabular}
\begin{tabular}{C{0.21\textwidth}|| C{0.21\textwidth}}
C & D
\end{tabular}
\begin{tabular}{|C{0.034\textwidth}|C{0.034\textwidth}|C{0.034\textwidth}|C{0.034\textwidth}||C{0.034\textwidth}|C{0.034\textwidth}|C{0.034\textwidth}|C{0.034\textwidth}|}
\hline
0 &0& 0.75 & 0.25 & 0.75 & 0.25 & 0 & 0\\
\hline
\end{tabular}
\end{table}

\noindent Using the first candidate $3N/2$ points shown in Fig.\ref{fig:p5}, the transmitted signal is:
\begin{table}[!h]
\centering
\label{parameters}
\renewcommand{\arraystretch}{1.0}
\begin{tabular}{C{0.137\textwidth}|| C{0.136\textwidth} || C{0.137\textwidth}}
\hline
$(A+C)$ & $(B+C)$ & $(B+D)$
\end{tabular}
\begin{tabular}{|C{0.016\textwidth}|C{0.016\textwidth}|C{0.016\textwidth}|C{0.016\textwidth}||C{0.016\textwidth}|C{0.016\textwidth}|C{0.016\textwidth}|C{0.016\textwidth}||C{0.016\textwidth}|C{0.016\textwidth}|C{0.016\textwidth}|C{0.016\textwidth}|}
\hline
0&0&1.75&0.25 & 1 & 0.71 & 0.75 & 0.25 & 1.75 & 0.96 & 0 & 0 \\
\hline
\end{tabular}
\end{table}

At the receiver, (A+D) is recovered as:
\begin{table}[!h]
\centering
\label{parameters}
\renewcommand{\arraystretch}{1.0}
\begin{tabular}{C{0.31\textwidth} }
\hline
$(A+D) = (A+C) +(B+D) - (B+C)$
\end{tabular}
\begin{tabular}{|C{0.06\textwidth}|C{0.06\textwidth}|C{0.06\textwidth}|C{0.06\textwidth}|}
\hline
0.75 & 0.25 & 1 & 0 \\
\hline
\end{tabular}
\end{table}
\\
Then $L$ and $R$  shown in Fig.\ref{fig:p1} are constructed, followed by the FFT block to recover the $N$ sub-carriers as:
\begin{table}[!h]
\centering
\renewcommand{\arraystretch}{1.0}
\begin{tabular}{C{0.21\textwidth}|| C{0.21\textwidth}}
\hline
$L = (A+C)-(B+D)$ & $R = (B+C) - (A+D)$
\end{tabular}
\begin{tabular}{|C{0.034\textwidth}|C{0.034\textwidth}|C{0.034\textwidth}|C{0.034\textwidth}||C{0.034\textwidth}|C{0.034\textwidth}|C{0.034\textwidth}|C{0.034\textwidth}|}
\hline
-1.75 &-0.96& 1.75 & 0.25 & 0.25 & 0.46 & -0.25 & 0.25\\
\hline
\end{tabular}
\begin{tabular}{C{0.447\textwidth}}
After the FFT block to recover $[X_0, X_1, ..., X_{7}]$
\end{tabular}
\begin{tabular}{|C{0.035\textwidth}|C{0.035\textwidth}|C{0.035\textwidth}|C{0.035\textwidth}|C{0.035\textwidth}|C{0.035\textwidth}|C{0.035\textwidth}|C{0.035\textwidth}|}
\hline
0 &-3-j & -3+j & -1+3j & 0 & -1-3j & -3-j & -3+j\\
\hline
\end{tabular}
\end{table}
\newpage
\noindent It could be seen that the messages carried by all sub-carriers are recovered.

\section{The Second Type of X-OFDM Waveform}\label{section22}
The second type of X-OFDM waveform is proposed to further increase the spectral efficiency at the sacrifice of the computational complexity, which are discussed in Section \ref{section3}. This section studies the system architecture and the signal processing at the transmitter and the receiver of the second type of X-OFDM waveform.

\subsection{System Architecture}
The system architecture of the second type of X-OFDM waveform is shown in Fig. \ref{SYSTEMMODEL2}. At the transmitter, there exist a splitter, two IFFT blocks, and a mixer. The splitter splits the $N$ sub-carriers into the odd sub-carriers and the even sub-carriers. The IFFT blocks transform the signal into the time domain, followed by the mixer to generate a $3N/2$ points signal. We note that the mixer is different to that of the first type of X-OFDM waveform. At the receiver, there exist a splitter, a recover, a mixer, a combiner, a IFFT and two FFT blocks. The splitter splits the $3N/2$ points into a $N$ points signal and a $N/2$ points signal. The FFT and recover blocks recover the odd sub-carriers. Then the mixer, combiner, IFFT and FFT blocks utilize the odd sub-carriers to successively recover the even sub-carriers.

\subsection{Signal Processing at Transmitter}
Similarly, the HS character of the sub-carriers in the frequency domain $[X_0, X_1, ..., X_{N-1}]$ guarantees the signal in the time domain to be real. The splitter splits the $N$ sub-carriers into the odd sub-carriers as $[0, X_1, 0, X_3, ..., X_{N-1}]$ and the even sub-carriers as $[X_0, 0, X_2, 0, ..., X_{N-2}, 0]$. After the two IFFT blocks, the signal in the time domain is obtained as Fig.\ref{fig:p2} and Fig.\ref{fig:p3}.

\begin{figure}[]
\centering
\includegraphics[width=0.5\textwidth]{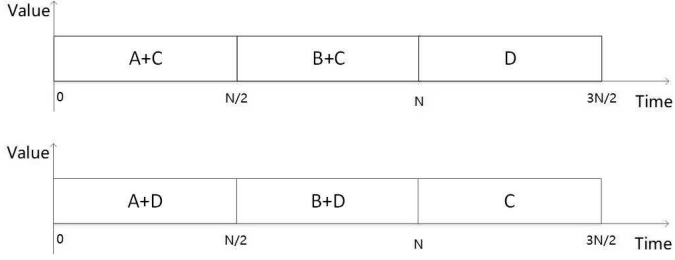}
\caption{The candidate transmitted signal in the time domain of the second type of  X-OFDM waveform.}
\label{fig:p6}
\end{figure}

Then we could transmit either (A + C), (B +
C), D  or  (A + D), (B + D), C as shown in Fig. \ref{fig:p6}.  The function of the mixer block is summarized as:
\begin{itemize}
\item  The $A$, $B$, $C$, $D$ are extracted using the equations (\ref{A}), (\ref{B}), (\ref{C}) and (\ref{D});
\item The $A$, $B$, $C$, $D$ are mixed to construct either $(A+C)$, $(B+C)$, $D$ or   $(A+D)$, $(B+D)$, $C$  as shown in Fig.\ref{fig:p6}.
\end{itemize}
Then the constructed 3N/2 points are transmitted at the transmitter.

\subsection{Signal Processing at Receiver}
The receiver receives the 3N/2 points of either $(A+C)$, $(B+C)$, $D$ or   $(A+D)$, $(B+D)$, $C$ as shown in Fig.\ref{fig:p6}. In the following, the $3N/2$ points $(A+C)$, $(B+C)$, $D$ are analyzed to recover the sub-carriers, and the $3N/2$ points $(A+D)$, $(B+D)$, $C$ follow the same way and are omitted to save space.

The splitter splits the $3N/2$ points  $(A+C)$, $(B+C)$, $D$  into the $N$ points  $(A+C)$, $(B+C)$ and the $N/2$ points $D$. Based on the  $N$ points  $(A+C)$, $(B+C)$, the odd sub-carriers are recovered by the FFT and recover blocks.  Assuming the signal after the FFT block is $[Y_0, Y_1, ..., Y_{N-1}]$ , we have:
\begin{subequations}
\begin{alignat}{3}
&Y_{2m+1} = \sum_{k=0}^{\frac{N}{2}-1}\left(x_k^A + x_k^C\right)\exp\left(-j\frac{2\pi}{N}(2m+1)k\right) + \notag \\
&+ \sum_{k=0}^{\frac{N}{2}-1}\left(x_{k}^B + x_{k}^C\right)\exp\left(-j\frac{2\pi}{N}(2m+1)\left(k + \frac{N}{2}\right)\right)\label{f_a}\\
&= \sum_{k=0}^{\frac{N}{2}-1}\left(\left(x_k^A + x_k^C\right)- \left(x_k^B + x_k^C\right)\right)\exp\left(-j\frac{2\pi}{N}(2m+1)k\right) \label{f_b}\\
&= \sum_{k=0}^{\frac{N}{2}-1}\left(x_k^A - x_k^B \right)\exp\left(-j\frac{2\pi}{N}(2m+1)k\right) \label{f_c}\\
&= \frac{1}{2}\left(\sum_{k=0}^{\frac{N}{2}-1}\left(x_k^A - x_k^B \right)\exp\left(-j\frac{2\pi}{N}(2m+1)k\right) \right.\notag \\
&\left. + \sum_{k=\frac{N}{2}}^{N-1}\left(x_k^B - x_k^A \right)\exp\left(-j\frac{2\pi}{N}(2m+1)k\right)\right)\label{f_d}\\
&= \frac{X_{2m+1}}{2}, 0 \leq m < \frac{N}{2}\label{f_e}
\end{alignat}
\end{subequations}\label{half_relation}

\noindent where equation (\ref{f_a}) uses the $N$ points  $(A+C)$, $(B+C)$ and equation (\ref{f_d}) to equation (\ref{f_e}) utilizes Fig. \ref{fig:p2} and the inverse of equations (\ref{AB_IFFT}), (\ref{A}) and (\ref{B}).  The recover block selects the odd parts multiplying by two to recover the odd sub-carriers $[0, X_1, 0, X_3, ..., X_{N-1}]$.

Then the even sub-carriers $[X_0, 0, X_2, 0, ..., X_{N-2}, 0]$ are recovered by the IFFT, mixer, combiner and FFT blocks. The IFFT block transforms the odd sub-carriers to the signal in the time domain as shown in Fig.\ref{fig:p2}. The function of the mixer block is summarized as:
\begin{itemize}
\item The $A$, $B$ are extracted using equations (\ref{A}), (\ref{B});
\item The $A$, $B$ and $(A+C)$, $(B+C)$ are mixed to obtain $C$.
\end{itemize}
The $N/2$ points $C$ could be obtained by $(A+C) - A$, $(B+C) - B$ or $\frac{(A+C)-A  + (B+C) - B}{2}$, where the third one uses the average to reduce the noise. Based on the $N/2$ points $C$ from the mixer block and the $N/2$ points $D$ from the splitter block, the combiner block constructs $(C-D)$, $(C-D)$ as shown in Fig.\ref{fig:p3}. The FFT block transforms the signal in the time domain to recover the odd sub-carriers as $[X_0, 0, X_2, 0, ..., X_{N-2}, 0]$.


\subsection{An Example}
Similarly, a simple example is provided for the second type of X-OFDM waveform.
At the transmitter, the sub-carrier number is $N = 8$ and the constellation is 16-QAM. The signal of the sub-carriers are:
\begin{table}[!h]
\centering
\renewcommand{\arraystretch}{1.0}
\begin{tabular}{C{0.447\textwidth}}
\hline
$N=8$ sub-carriers $[X_0, X_1, ..., X_{7}]$ satisfying the HS requirement
\end{tabular}
\begin{tabular}{|C{0.035\textwidth}|C{0.035\textwidth}|C{0.035\textwidth}|C{0.035\textwidth}|C{0.035\textwidth}|C{0.035\textwidth}|C{0.035\textwidth}|C{0.035\textwidth}|}
\hline
0 &-3-j & -3+j & -1+3j & 0 & -1-3j & -3-j & -3+j\\
\hline
\end{tabular}
\end{table}

\noindent The odd sub-carriers and $A$, $B$ are:
\begin{table}[!h]
\centering
\renewcommand{\arraystretch}{1.0}
\begin{tabular}{C{0.447\textwidth}}
\hline
The odd sub-carriers $[0, X_1, 0, X_3, ..., 0, X_{7}]$
\end{tabular}
\begin{tabular}{|C{0.035\textwidth}|C{0.035\textwidth}|C{0.035\textwidth}|C{0.035\textwidth}|C{0.035\textwidth}|C{0.035\textwidth}|C{0.035\textwidth}|C{0.035\textwidth}|}
\hline
0 &-3-j & 0 & -1+3j & 0 & -1-3j & 0 & -3+j\\
\hline
\end{tabular}
\begin{tabular}{C{0.447\textwidth}}
After the IFFT block to get $[x_0^{odd}, x_1^{odd}, ..., x_7^{odd}]$
\end{tabular}
\begin{tabular}{|C{0.035\textwidth}|C{0.035\textwidth}|C{0.035\textwidth}|C{0.035\textwidth}|C{0.035\textwidth}|C{0.035\textwidth}|C{0.035\textwidth}|C{0.035\textwidth}|}
\hline
-1 &-0.71& 1 & 0 & 1 & 0.71 & -1 & 0\\
\hline
\end{tabular}
\begin{tabular}{C{0.21\textwidth}|| C{0.21\textwidth}}
A & B
\end{tabular}
\begin{tabular}{|C{0.034\textwidth}|C{0.034\textwidth}|C{0.034\textwidth}|C{0.034\textwidth}||C{0.034\textwidth}|C{0.034\textwidth}|C{0.034\textwidth}|C{0.034\textwidth}|}
\hline
0 &0& 1 & 0 & 1 & 0.71 & 0 & 0\\
\hline
\end{tabular}
\end{table}
\\
The even sub-carriers and $C$, $D$ are:
\begin{table}[!h]
\centering
\renewcommand{\arraystretch}{1.0}
\begin{tabular}{C{0.447\textwidth}}
\hline
The even sub-carriers $[X_0, 0, X_2, 0, ..., X_6, 0]$
\end{tabular}
\begin{tabular}{|C{0.035\textwidth}|C{0.035\textwidth}|C{0.035\textwidth}|C{0.035\textwidth}|C{0.035\textwidth}|C{0.035\textwidth}|C{0.035\textwidth}|C{0.035\textwidth}|}
\hline
0 &0 & -3+j & 0 & 0 & 0 & -3-j & 0\\
\hline
\end{tabular}
\begin{tabular}{C{0.447\textwidth}}
After the IFFT block to get $[x_0^{even}, x_1^{even}, ..., x_7^{even}]$
\end{tabular}
\begin{tabular}{|C{0.035\textwidth}|C{0.035\textwidth}|C{0.035\textwidth}|C{0.035\textwidth}|C{0.035\textwidth}|C{0.035\textwidth}|C{0.035\textwidth}|C{0.035\textwidth}|}
\hline
-0.75 &-0.25& 0.75 & 0.25 & -0.75 & -0.25 & 0.75 & 0.25\\
\hline
\end{tabular}
\begin{tabular}{C{0.21\textwidth}|| C{0.21\textwidth}}
C & D
\end{tabular}
\begin{tabular}{|C{0.034\textwidth}|C{0.034\textwidth}|C{0.034\textwidth}|C{0.034\textwidth}||C{0.034\textwidth}|C{0.034\textwidth}|C{0.034\textwidth}|C{0.034\textwidth}|}
\hline
0 &0& 0.75 & 0.25 & 0.75 & 0.25 & 0 & 0\\
\hline
\end{tabular}
\end{table}
\\
Using the first candidate $3N/2$ points shown in Fig.\ref{fig:p6}, the transmitted signal is:
\begin{table}[!h]
\centering
\renewcommand{\arraystretch}{1.0}
\begin{tabular}{C{0.137\textwidth}|| C{0.136\textwidth} || C{0.137\textwidth}}
\hline
$(A+C)$ & $(B+C)$ & $D$
\end{tabular}
\begin{tabular}{|C{0.016\textwidth}|C{0.016\textwidth}|C{0.016\textwidth}|C{0.016\textwidth}||C{0.016\textwidth}|C{0.016\textwidth}|C{0.016\textwidth}|C{0.016\textwidth}||C{0.016\textwidth}|C{0.016\textwidth}|C{0.016\textwidth}|C{0.016\textwidth}|}
\hline
0&0&1.75&0.25 & 1 & 0.71 & 0.75 & 0.25 & 0.75 & 0.25 & 0 & 0 \\
\hline
\end{tabular}
\end{table}

At the receiver, the input signal is split as:
\begin{table}[!h]
\centering
\renewcommand{\arraystretch}{1.0}
\begin{tabular}{C{0.137\textwidth}|| C{0.136\textwidth}}
\hline
$(A+C)$ & $(B+C)$
\end{tabular}
\begin{tabular}{|C{0.016\textwidth}|C{0.016\textwidth}|C{0.016\textwidth}|C{0.016\textwidth}||C{0.016\textwidth}|C{0.016\textwidth}|C{0.016\textwidth}|C{0.016\textwidth}|}
\hline
0&0&1.75&0.25 & 1 & 0.71 & 0.75 & 0.25 \\
\hline
\end{tabular}
\end{table}
\begin{table}[!h]
\centering
\renewcommand{\arraystretch}{1.0}
\begin{tabular}{C{0.137\textwidth}}
\hline
 $D$
\end{tabular}\\
\begin{tabular}{|C{0.016\textwidth}|C{0.016\textwidth}|C{0.016\textwidth}|C{0.016\textwidth}|}
\hline
 0.75 & 0.25 & 0 & 0 \\
\hline
\end{tabular}
\end{table}

\noindent The odd sub-carriers are recovered as:
\begin{table}[!h]
\centering
\renewcommand{\arraystretch}{1.0}
\begin{tabular}{C{0.21\textwidth}|| C{0.21\textwidth}}
\hline
$(A+C)$ & $(B+C)$
\end{tabular}
\begin{tabular}{|C{0.034\textwidth}|C{0.034\textwidth}|C{0.034\textwidth}|C{0.034\textwidth}||C{0.034\textwidth}|C{0.034\textwidth}|C{0.034\textwidth}|C{0.034\textwidth}|}
\hline
0&0&1.75&0.25 & 1 & 0.71 & 0.75 & 0.25 \\
\hline
\end{tabular}
\begin{tabular}{C{0.447\textwidth}}
After the FFT block to get $[Y_0, Y_1, ..., Y_{7}]$
\end{tabular}
\begin{tabular}{|C{0.034\textwidth}|C{0.034\textwidth}|C{0.034\textwidth}|C{0.034\textwidth}||C{0.034\textwidth}|C{0.034\textwidth}|C{0.034\textwidth}|C{0.034\textwidth}|}
\hline
4.71 &-1.5 - 0.5j & -1.5 -0.2j & -0.5 +1.5j & 2.29 & -0.5 -1.5j & -1.5 +0.2j & -1.5 +0.5j\\
\hline
\end{tabular}
\begin{tabular}{C{0.447\textwidth}}
After the recover block to get \\$[0, X_1, 0, X_3 ,..., 0, X_{7}]$ = $2\times [0, Y_1, 0, Y_3 ,..., 0, Y_{7}]$
\end{tabular}
\begin{tabular}{|C{0.034\textwidth}|C{0.034\textwidth}|C{0.034\textwidth}|C{0.034\textwidth}||C{0.034\textwidth}|C{0.034\textwidth}|C{0.034\textwidth}|C{0.034\textwidth}|}
\hline
0 &-3-j & 0 & -1+3j & 0 & -1-3j & 0 & -3+j\\
\hline
\end{tabular}
\end{table}

\newpage
\noindent The even sub-carriers are recovered by four steps. The first step re-extracts $A$, $B$ as:
\begin{table}[!h]
\centering
\label{parameters}
\renewcommand{\arraystretch}{1.0}
\begin{tabular}{C{0.447\textwidth}}
\hline
Odd sub-carriers $[0, X_1, 0, X_3, ..., 0, X_{7}]$
\end{tabular}
\begin{tabular}{|C{0.035\textwidth}|C{0.035\textwidth}|C{0.035\textwidth}|C{0.035\textwidth}|C{0.035\textwidth}|C{0.035\textwidth}|C{0.035\textwidth}|C{0.035\textwidth}|}
\hline
0 &-3-j & 0 & -1+3j & 0 & -1-3j & 0 & -3+j\\
\hline
\end{tabular}
\begin{tabular}{C{0.447\textwidth}}
After IFFT to get $[x_0^{odd}, x_1^{odd}, ..., x_7^{odd}]$
\end{tabular}
\begin{tabular}{|C{0.035\textwidth}|C{0.035\textwidth}|C{0.035\textwidth}|C{0.035\textwidth}|C{0.035\textwidth}|C{0.035\textwidth}|C{0.035\textwidth}|C{0.035\textwidth}|}
\hline
-1 &-0.71& 1 & 0 & 1 & 0.71 & -1 & 0\\
\hline
\end{tabular}
\begin{tabular}{C{0.21\textwidth}|| C{0.21\textwidth}}
A & B
\end{tabular}
\begin{tabular}{|C{0.034\textwidth}|C{0.034\textwidth}|C{0.034\textwidth}|C{0.034\textwidth}||C{0.034\textwidth}|C{0.034\textwidth}|C{0.034\textwidth}|C{0.034\textwidth}|}
\hline
0 &0& 1 & 0 & 1 & 0.71 & 0 & 0\\
\hline
\end{tabular}
\end{table}
\\
\noindent The second step re-constructs $C$ as:
\begin{table}[!h]
\centering
\renewcommand{\arraystretch}{1.0}
\begin{tabular}{C{0.39\textwidth}}
\hline
$C = (A+C) - A  = (B+C) - B
  = \frac{(A+C) - A + (B+C) - B}{2} $
\end{tabular}\\
\begin{tabular}{|C{0.08\textwidth}|C{0.08\textwidth}|C{0.08\textwidth}|C{0.08\textwidth}|}
\hline
 0 &0 & 0.75 & 0.25 \\
\hline
\end{tabular}
\end{table}
\\
The third step combines $C$, $D$ as:
\begin{table}[!h]
\centering
\renewcommand{\arraystretch}{1.0}
\begin{tabular}{C{0.21\textwidth}|| C{0.21\textwidth}}
\hline
$(C-D)$ & $(C-D)$
\end{tabular}
\begin{tabular}{|C{0.035\textwidth}|C{0.035\textwidth}|C{0.035\textwidth}|C{0.035\textwidth}|C{0.035\textwidth}|C{0.035\textwidth}|C{0.035\textwidth}|C{0.035\textwidth}|}
\hline
-0.75 & -0.25&0.75&0.25 & -0.75&-0.25&0.75&0.25 \\
\hline
\end{tabular}
\end{table}
\\The fourth step recovers the even sub-carriers as:
\begin{table}[!h]
\centering
\renewcommand{\arraystretch}{1.0}
\begin{tabular}{C{0.447\textwidth}}
\hline
After the FFT block to recover $[X_0, 0, X_2, 0, ..., X_{6}, 0]$
\end{tabular}
\begin{tabular}{|C{0.035\textwidth}|C{0.035\textwidth}|C{0.035\textwidth}|C{0.035\textwidth}|C{0.035\textwidth}|C{0.035\textwidth}|C{0.035\textwidth}|C{0.035\textwidth}|}
\hline
0 & 0& -3+j &0 & 0&0&-3-j&0 \\
\hline
\end{tabular}
\end{table}
\\
\noindent  It could be seen that the messages carried by the odd sub-carriers and the even sub-carriers are recovered successively.

\section{Discussions}\label{section3}
The spectral efficiency, computational complexity, motivation and comparison of the proposed two types of X-OFDM waveform are discussed in this section.

\subsection{Spectral Efficiency}
Due to the HS requirement, the second half sub-carriers is determined by the first half sub-carriers as equation (\ref{HS_equal}), at the sacrifice of $1/2$ spectral efficiency. In the time domain, the transmitted signal extends the $N$ sub-carriers to $3N/2$ points  as shown in Fig. \ref{fig:p5} and Fig. \ref{fig:p6}, at the sacrifice of $2/3$ spectral efficiency. Hence the spectral efficiency for the two types of X-OFDM waveform is:
\begin{align}
SE = 1/3 = 1/2 \times 2/3.
\end{align}
Moreover, the spectral efficiency is $1/2$ for the DCO-OFDM waveform, and $1/4$ for the ACO-OFDM waveform and the U-OFDM waveform \cite{book, ADO-OFDM}.

\subsection{Complexity}
The computational complexity is defined as the number of the multiplications and the additional operations at both the transmitter and the receiver. For an IFFT/FFT block with size of $N$, it requires approximately $N\log(N)$ operations \cite{complexity1, complexity2}. For the system architecture of the first type of X-OFDM waveform shown in Fig. \ref{SYSTEMMODEL}, it has two IFFT blocks at the transmitter and one FFT block at the receiver. The computational complexity is around $3N\log(N)$. For the system architecture of the second type of X-OFDM waveform shown in Fig. \ref{SYSTEMMODEL2}, it has two IFFT blocks at the transmitter and two FFT and one IFFT blocks at the receiver. The computational complexity is around $5N\log(N)$. Then the total computational complexity of the DCO-OFDM, the ACO-OFDM, the U-OFDM and the two types of X-OFDM waveform are summarized in TABLE \ref{compleixty_table}.

\begin{table}[!h]
\centering
\caption{The Computational Complexity of the DCO-OFDM, the ACO-OFDM, the U-OFDM and the two types of X-OFDM Waveform.}
\label{compleixty_table}
\renewcommand{\arraystretch}{1.0}
\begin{tabular}{p{0.09\textwidth}|p{0.09\textwidth}|p{0.09\textwidth}|p{0.1\textwidth}}
\hline
OFDM Waveforms &Number of IFFT block at transmitter& Number of FFT/IFFT block at receiver & Total computational complexity\\
\hline
DCO-OFDM &1&1& $\mathcal O (2N \log_2N)$\\
\hline
ACO-OFDM &1&1& $\mathcal O (2N \log_2N)$\\
\hline
U-OFDM &1&1& $\mathcal O (2N \log_2N)$\\
\hline
First type & & &
\\ of X-OFDM &2&1& $\mathcal O (3N \log_2N)$\\
\hline
Second type & & &
\\ of X-OFDM &2&3& $\mathcal O (5N \log_2N)$\\
\hline
\end{tabular}
\end{table}

\subsection{Motivation and Comparison}\label{mc}
The proposed two types of X-OFDM waveform utilize the antisymmetric and symmetric characters of the signal in the time domain,  when the odd and even sub-carriers in the frequency domain satisfy the HS requirement.

For the first type of X-OFDM waveform, based on the forward analysis with the IFFT block, the key motivation is the relationship as equation (\ref{ABCD}). Then either three of $(A + C)$, $(B + D)$, $(B + C)$, $(A + D)$ as shown in Fig.2e are transmitted, and the remaining one is recoverable. However, the recovering operation highly increase the noise. For an example with $(A + C)$, $(B + D)$, $(B + C)$ being transmitted, and the thermal noise with variance $\sigma^2$ at the receiver, we re-write equation (\ref{LR}) as:
\begin{align}
&L = (A+C) - (B+D) \notag \\
&R = 2(B+C) - (A+C) - (B+D)
\end{align}
The add and subtract operations increase the noise variance of $L$ and $R$ to be $2\sigma^2$ and $6\sigma^2$, which degrade the system data rate even that the noise of $L$ and $R$ are dependent.

For the second type of X-OFDM waveform, to decrease the influence of noise and based on the backward analysis with the FFT block, the key motivation is as follows:
\begin{itemize}
\item For the even sub-carriers satisfying the HS requirement, after the IFFT block, the signal in the time domain is symmetric as shown in Fig.\ref{fig:p3}.
\item Reversely, for the symmetric signal in the time domain, after the FFT block, the signal has values at the even sub-carriers, but equals $0$ at the odd sub-carriers.
\item Then for the $N$ points  signal $(A+C)$, $(B+C)$ or $(A+D)$, $(B+D)$ in the time domain, after the FFT block, the signal has useless values at the even sub-carriers, but recovers the odd sub-carriers.
\item Once the odd sub-carriers are recovered, the even sub-carriers could be recovered successively.
\end{itemize}
The procedures of the second type of X-OFDM waveform are complicated, but do not have the add and subtract operations, which could decrease the noise and enhance the system data rate.

Therefore, the first type X-OFDM waveform has the lower system data rate, but the lower computational complexity. The second type X-OFDM waveform has the higher system data rate, but the higher computational complexity.

\section{Numerical Simulations}\label{section4}
In this section, numerical simulations are provided to evaluate the performance of the proposed two types of X-OFDM waveform. 

\subsection{Definitions}

For the simulations, we consider the improved free-space optical intensity channel \cite{channel1,channel2,channel3} as:
\begin{align}
Y = X + \sqrt X Z_1 + Z_0
\end{align}
where X is the received signal generated by the DCO-OFDM,  the ACO-OFDM, the U-OFDM, or the two types of X-OFDM waveform, $Z_0$ is the signal-independent Gaussian noise with zero mean and $\sigma^2$ variance, and $Z_1$ is the signal-dependent Gaussian noise with zero mean and $\xi^2$ variance. The signal-independent noise mainly consists of the background radiation noise received by the photodiode (PD) and the thermal noise from the electrical circuits at the receiver \cite{int_noise1,int_noise2}. The signal-dependent noise mainly has the relative intensity noise from the laser diode (LD)  or light emitting diode (LED) \cite{RN1,RN2} at the transmitter and the shot noise from the PD \cite{SN1,SN2} at the receiver.

 The electrical power per bit to the signal-dependent noise ratio is defined as:
\begin{align}
<E_{b(elec)}/{\xi^2}> = \frac{E(X^2)}{\xi^2  \cdot b_{s}}.
\end{align}
where $b_{s}$ is the number of bits per symbol depending on the constellation and the spectral efficiency.

 The electrical power per bit to signal-independent noise ratio is defined as:
\begin{align}
<E_{b(elec)}/{\sigma^2}> = \frac{E(X^2)}{\sigma^2 \cdot b_{s}}\label{ref_snr}
\end{align}

Regardless of the electrical current threshold, the output optical power of the LD or LED increases linearly with the input electrical current \cite{RN1,RN2}. Hence the optical power per bit to signal-dependent noise ratio is defined as:
\begin{align}
<E_{b(opt)}/{\xi^2}> = \frac{E(X)}{\xi^2  \cdot b_{s}}.
\end{align}

The optical power per bit to signal-independent noise ratio is defined as:
\begin{align}
<E_{b(opt)}/{\sigma^2}> = \frac{E(X)}{\sigma^2  \cdot b_{s}}.
\end{align}

For the DCO-OFDM waveform, the DC bias level is set relative to the standard deviation of X. With the DC bias as:
\begin{align}
DC_{bias} = \mu \sqrt{E(X^2)},
\end{align}
the bias level is $10\log_{10}(\mu^2+1)$ dB \cite{ADO-OFDM}.

\begin{figure}[]
\centering
\includegraphics[width=0.5\textwidth]{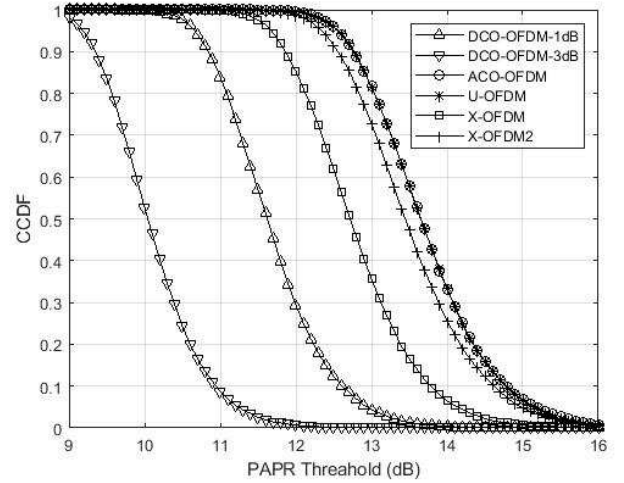}
\caption{The CCDF of the PAPR for the DCO-OFDM with 1dB and 3dB DC bias level, the ACO-OFDM, the U-OFDM and the two types of X-OFDM waveform.}
\label{PAPR}
\end{figure}

\subsection{Simulations}
Using Monte Carlo simulation, the results are the average of $2000$ OFDM symbols with $2048$ sub-carriers, for different constellation, waveform, bit error rate (BER), $<E_{b(elec)}/{\xi^2}>$, $<E_{b(opt)}/{\xi^2}>$, $<E_{b(elec)}/{\sigma^2}>$,  and $<E_{b(opt)}/{\sigma^2}>$.

\subsubsection{PAPR Performance}
The complementary cumulative distribution function (CCDF) of the PAPR for the DCO-OFDM with 1dB and 3dB DC bias level, the ACO-OFDM, the U-OFDM and the two types of X-OFDM waveform are shown in Fig. \ref{PAPR}. It could be seen that the PAPR performance of the ACO-OFDM waveform and that of the U-OFDM waveform overlap each other. The DCO-OFDM waveform with the DC bias outperforms the ACO-OFDM/U-OFDM and the two types of X-OFDM waveform without the DC bias. Meanwhile, the PAPR performance gets better as the DC bias level of the DCO-OFDM waveform increases. The reason is that, for the DCO-OFDM waveform, both the peak power and the average power increase with the DC bias level. For the scenarios without the DC bias, the first type of X-OFDM waveform has the best PAPR performance. The second type of X-OFDM waveform has better PAPR performance than that of the ACO-OFDM/U-OFDM waveform.

\begin{figure}[]
\centering
\includegraphics[width=0.5\textwidth]{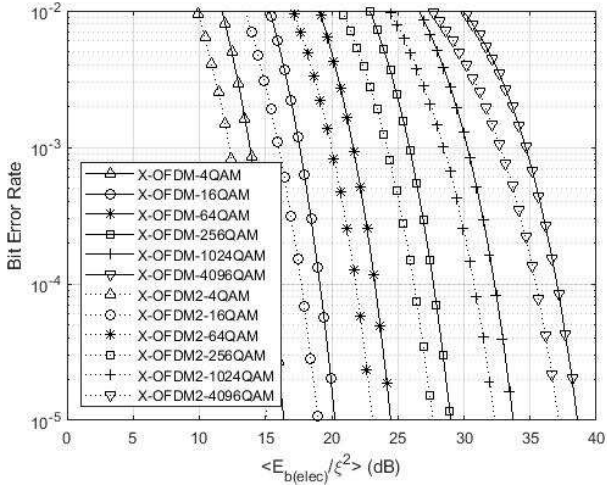}
\caption{The BER against the $<E_{b(elec)}/{\xi^2}> $ for the two types of X-OFDM waveform, when only the signal-dependent noise exists.}
\label{BER1}
\end{figure}

\begin{figure}[]
\centering
\includegraphics[width=0.5\textwidth]{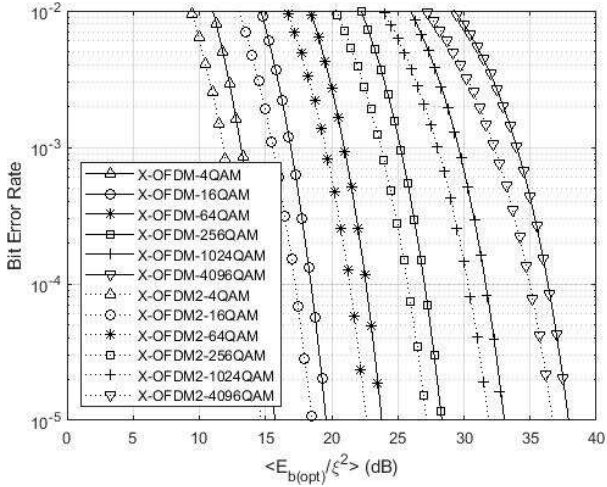}
\caption{The BER against the $<E_{b(opt)}/{\xi^2}> $ for the two types of X-OFDM waveform, when only the signal-dependent noise exists.}
\label{BER2}
\end{figure}

\subsubsection{BER Performance}
The BER performance of the two types of X-OFDM waveform are provided for the scenario with only the signal-dependent noise, and the other scenarios are compared in next subsection.

For the two types of X-OFDM, when only the signal-independent noise exists, the BER against the $<E_{b(elec)}/{\xi^2}>$ and the $<E_{b(opt)}/{\xi^2}>$  are shown in Fig.\ref{BER1} and Fig.\ref{BER2}. It is observed that the second type of X-OFDM waveform has better BER performance with the same $<E_{b(elec)}/{\xi^2}>$ or $<E_{b(opt)}/{\xi^2}>$. Since the decoding is processed in the electrical domain, the analysis in Section \ref{mc} explains the reason that the second type of X-OFDM waveform has better BER performance with the same $<E_{b(elec)}/{\xi^2}>$. The simulation results show that this character also holds for the $<E_{b(opt)}/{\xi^2}>$. 

\begin{figure}[]
\centering
\includegraphics[width=0.5\textwidth]{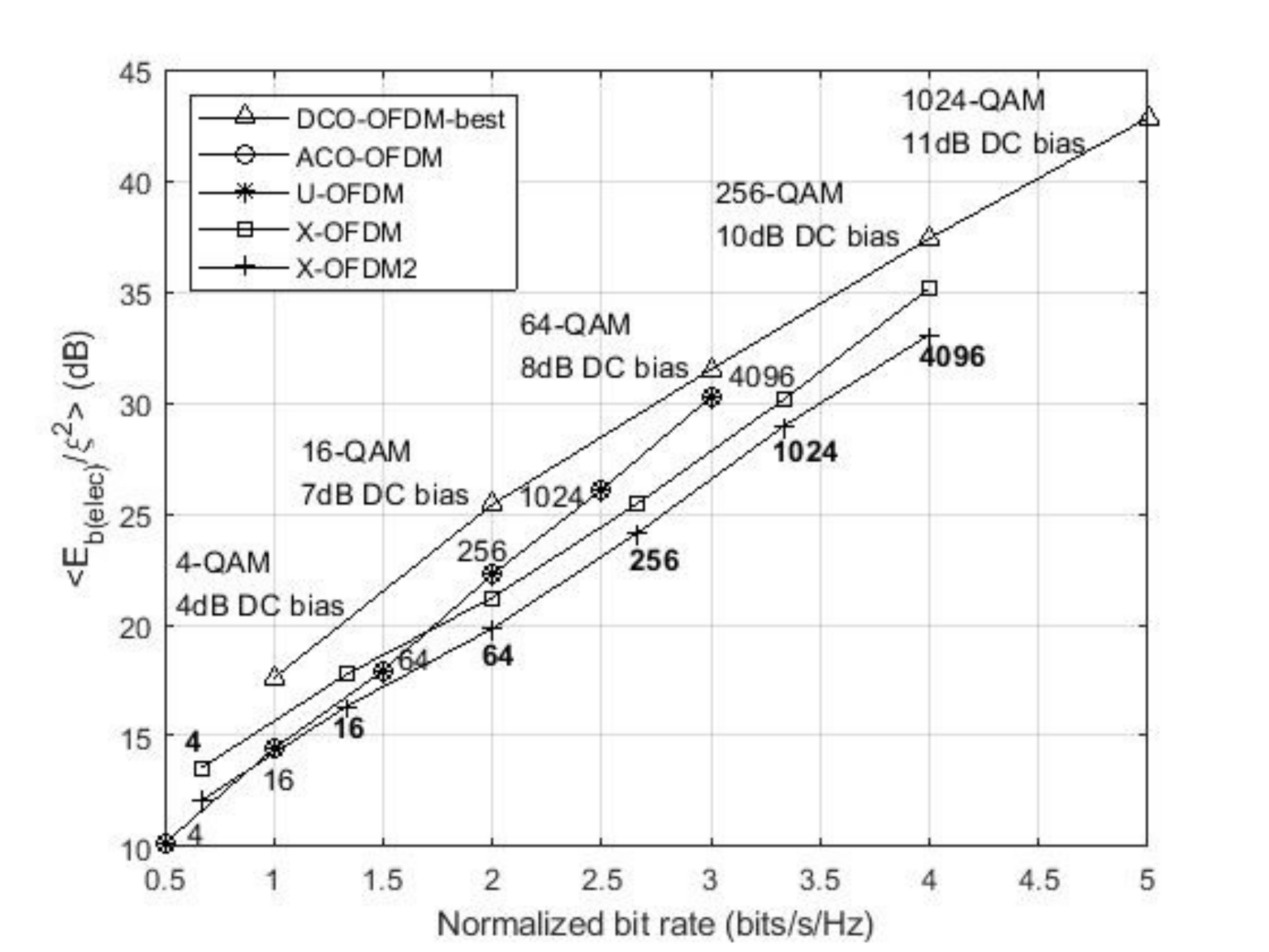}
\caption{The $<E_{b(elec)}/{\xi^2}> $ against the normalized bit rate for the DCO-OFDM with the optimal DC bias level, the ACO-OFDM, the U-OFDM and the two types of X-OFDM waveform, when the BER is $10^{-3}$ and only the signal-dependent noise exists.}
\label{P1_dep}
\end{figure}

\begin{figure}[]
\centering
\includegraphics[width=0.5\textwidth]{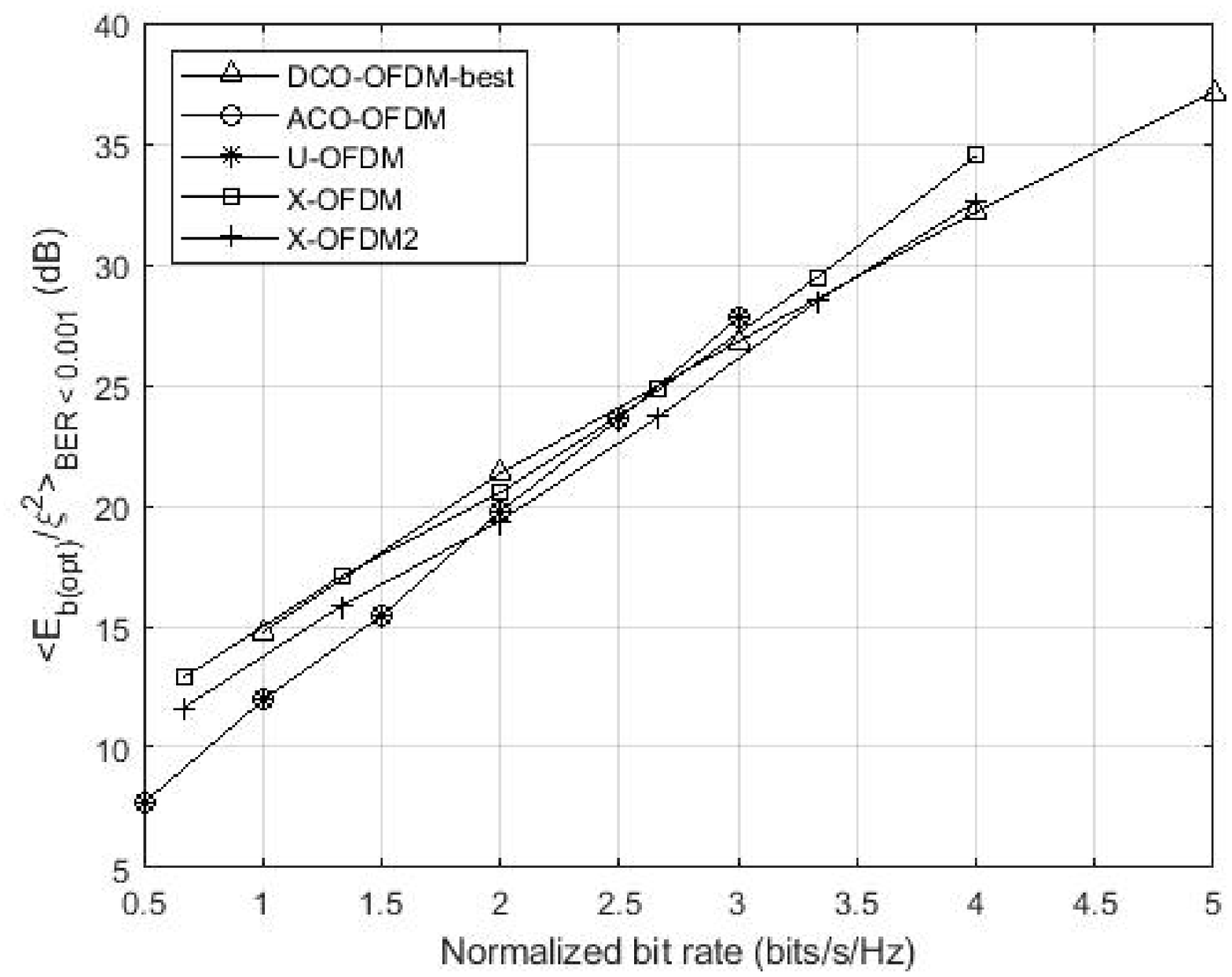}
\caption{The $<E_{b(opt)}/{\xi^2}> $ against the normalized bit rate for the DCO-OFDM with the optimal DC bias level, the ACO-OFDM, the U-OFDM and the two types of X-OFDM waveform, when the BER is $10^{-3}$ and only the signal-dependent noise exists.}
\label{P2_dep}
\end{figure}

%
%

\begin{figure}[]
\centering
\includegraphics[width=0.5\textwidth]{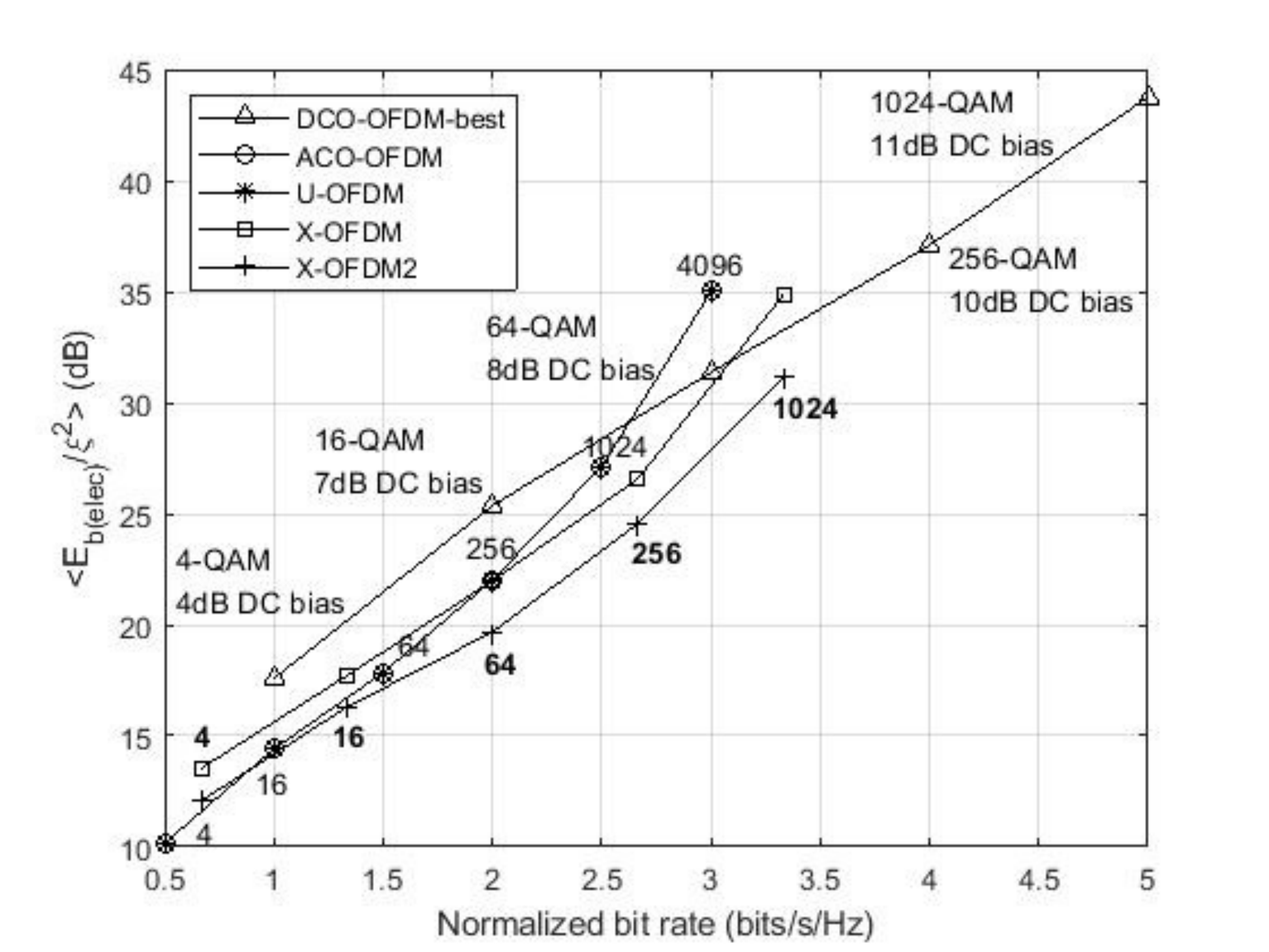}
\caption{The $<E_{b(elec)}/{\xi^2}> $ against the normalized bit rate for the DCO-OFDM with the optimal DC bias level, the ACO-OFDM, the U-OFDM and the two types of X-OFDM waveform, when the BER is $10^{-3}$, and the signal-dependent noise exists and the signal-independent noise is set to be $E(X^2)/\sigma^2 = 40 \text{ dB}$.}
\label{P1_dep_40}
\end{figure}

\begin{figure}[]
\centering
\includegraphics[width=0.5\textwidth]{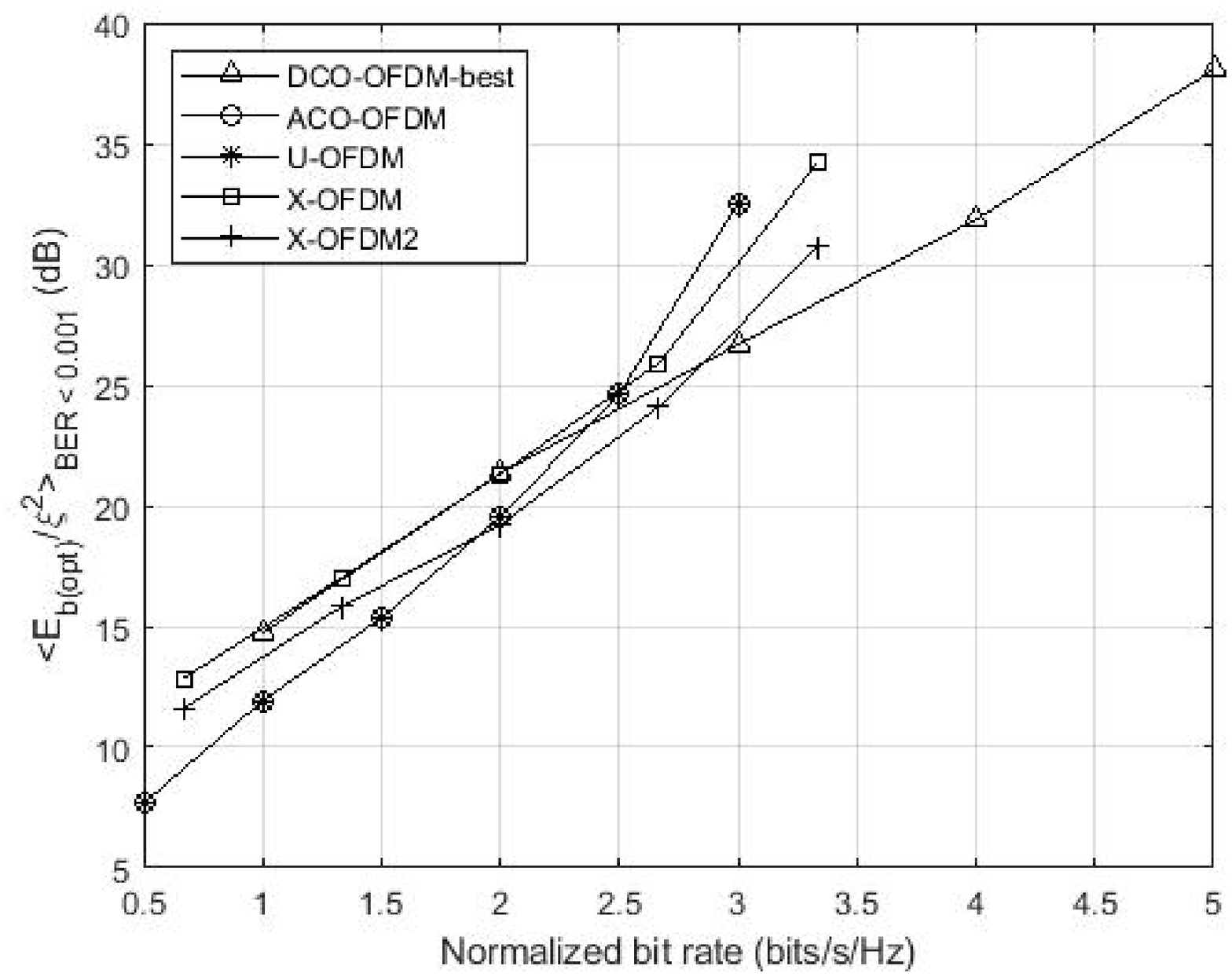}
\caption{The $<E_{b(opt)}/{\xi^2}> $ against the normalized bit rate for the DCO-OFDM with the optimal DC bias level, the ACO-OFDM, the U-OFDM and the two types of X-OFDM waveform, when the BER is $10^{-3}$, and the signal-dependent noise exists and the signal-independent noise is set to be $E(X^2)/\sigma^2 = 40 \text{ dB}$.}
\label{P2_dep_40}
\end{figure}

\begin{figure}[]
\centering
\includegraphics[width=0.5\textwidth]{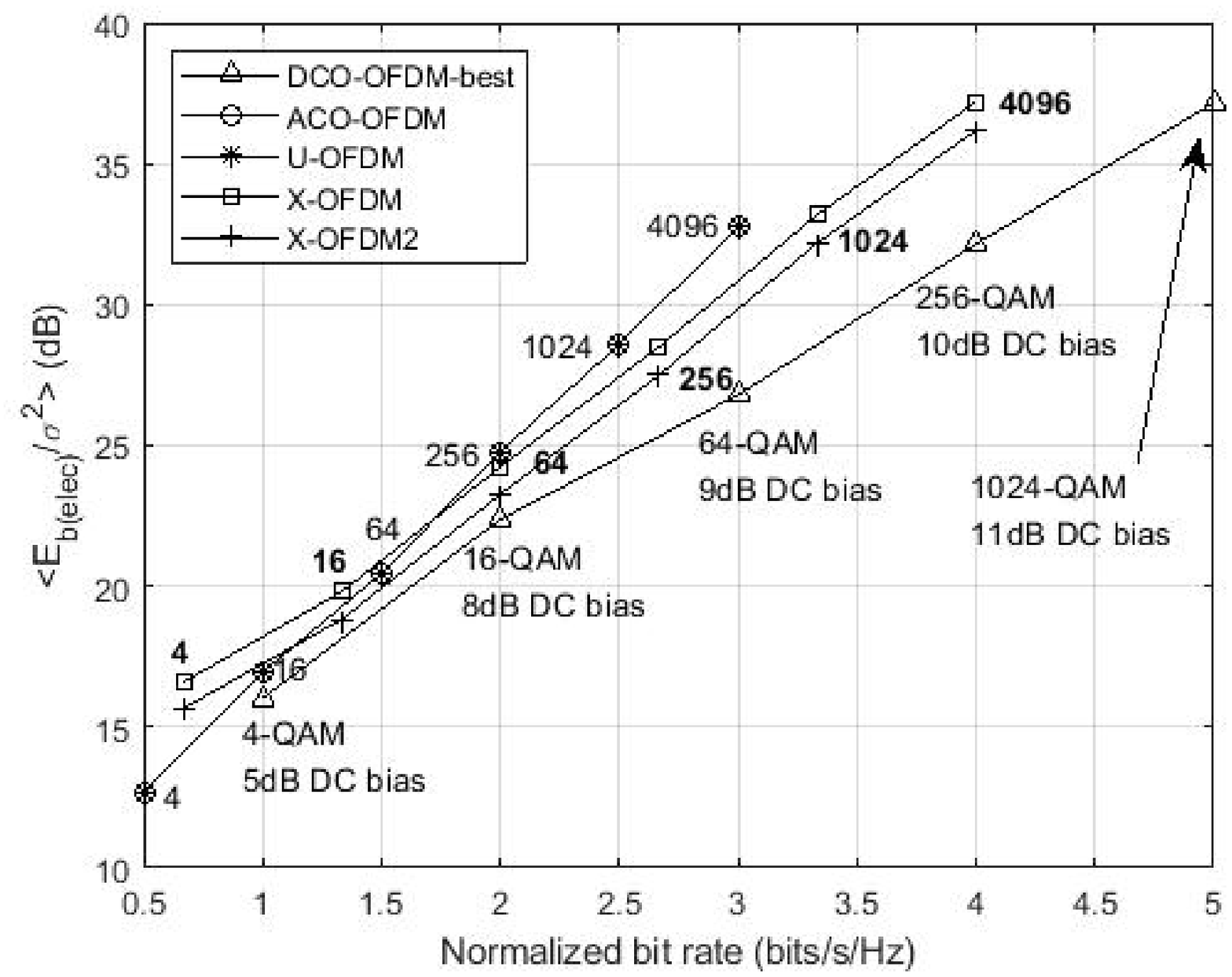}
\caption{The $<E_{b(elec)}/{\sigma^2}> $ against the normalized bit rate for the DCO-OFDM with the optimal DC bias level, the ACO-OFDM, the U-OFDM and the two types of X-OFDM waveform, when the BER is $10^{-3}$ and only the signal-independent noise exists.}
\label{P1_gauss}
\end{figure}

\begin{figure}[]
\centering
\includegraphics[width=0.5\textwidth]{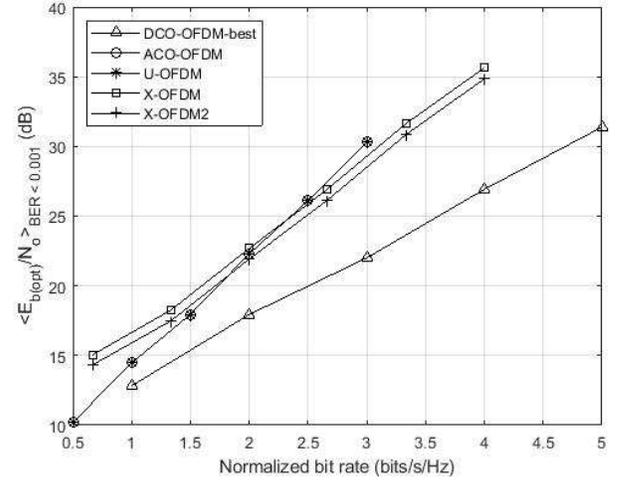}
\caption{The $<E_{b(opt)}/{\sigma^2}> $ against the normalized bit rate for the DCO-OFDM with the optimal DC bias level, the ACO-OFDM, the U-OFDM and the two types of X-OFDM waveform, when the BER is $10^{-3}$ and only the signal-independent noise exists.}
\label{P2_gauss}
\end{figure}

\subsubsection{Performance Comparison with the BER of $10^{-3}$ } The required  $<E_{b(elec)}/{\xi^2}>$, $<E_{b(opt)}/{\xi^2}>$, $<E_{b(elec)}/{\sigma^2}>$,  and $<E_{b(opt)}/{\sigma^2}>$ for the BER of $10^{-3}$ is used for the comparison between the different waveform. During the simulations, the DCO-OFDM waveform with the DC bias level $\{1, 2, 3, 4, 5, 6, 7$, $8, 9, 10, 11, 12, 13\}$ dB are simulated, and the optimal ones are selected as the result of the DCO-OFDM waveform with the best DC bias level. Meanwhile, the normalized bit rate of the DCO-OFDM, the ACO-OFDM, the U-OFDM  and the two types of X-OFDM waveform are $1/2, 1/4, 1/4$ \cite{book, ADO-OFDM} and $1/3$  spectral efficiency times the constellation.


For the scenario with only the signal-dependent noise, the best DC bias level for the DCO-OFDM waveform are $\{4, 7, 8, 10, 11\}$ dB  for $\{4, 16, 64, 256, 1024\}$ QAM constellation. When the BER is $10^{-3}$, the $<E_{b(elec)}/{\xi^2}>$ and the $<E_{b(opt)}/{\xi^2}>$ against the normalized bit rate  for  the different waveform are shown in Fig.\ref{P1_dep} and Fig.\ref{P2_dep}. It could be seen that the performance of the ACO-OFDM waveform and that of the U-OFDM waveform overlap each other. Meanwhile, the second type of X-OFDM waveform is better than the first type of X-OFDM waveform as discussed before. For the  $<E_{b(elec)}/{\xi^2}>$ in Fig.\ref{P1_dep}, the DCO-OFDM waveform with the best DC bias level has the worst power efficiency.  The second type of X-OFDM waveform outperforms the ACO-OFDM/U-OFDM waveform and is the best waveform,  when the normalized bit rate is larger than 0.9 bits/s/Hz. The first type of X-OFDM waveform outperforms the ACO-OFDM/U-OFDM waveform, when the normalized bit rate is larger than 1.7 bits/s/Hz. For the  $<E_{b(opt)}/{\xi^2}>$ in Fig.\ref{P2_dep}, the best waveform is the ACO-OFDM/U-OFDM waveform, the second type of X-OFDM waveform, and the DCO-OFDM waveform with the best DC bias level, when the normalized bit rate is smaller than 1.88, in [1.88, 3.46], and larger than 3.46 bits/s/Hz respectively.


For the scenario where the signal-dependent noise exists and the signal-independent noise is set to be $<E(X^2)/{\sigma^2}> = 40 \text{ dB}$, the best DC bias level are also $\{4, 7, 8, 10, 11\}$ dB  for $\{4, 16, 64, 256, 1024\}$  QAM constellation. When the BER is $10^{-3}$, the $<E_{b(elec)}/{\xi^2}>$ and the $<E_{b(opt)}/{\xi^2}>$ against the normalized bit rate  for  the different waveform are shown in Fig.\ref{P1_dep_40} and Fig.\ref{P2_dep_40}. It could be seen that the performance of the ACO-OFDM waveform and that of the U-OFDM waveform overlap each other. Meanwhile, the two types of X-OFDM waveform with the 4096-QAM could not achieve the BER of $10^3$ with  $<E(X^2)/{\sigma^2}> = 40 \text{dB}$, which will be discussed together with the scenario with only the signal-independent noise later. For the  $<E_{b(elec)}/{\xi^2}>$ in Fig.\ref{P1_dep_40}, the best waveform is the ACO-OFDM/U-OFDM waveform, the second type of X-OFDM waveform, when the normalized bit rate is smaller and larger than 0.9 bit/s/Hz. For the  $<E_{b(opt)}/{\xi^2}>$ in Fig.\ref{P2_dep_40}, the best waveform is the ACO-OFDM/U-OFDM waveform, the second type of X-OFDM waveform, and the DCO-OFDM waveform with the best DC bias level, when the normalized bit rate is smaller than 1.91, in [1.91, 2.84], and larger than 2.84 bits/s/Hz respectively.

For the scenario with only the signal-independent noise, the best DC bias level are  $\{5, 8, 9, 10, 11\}$ dB  for $\{4, 16, 64, 256, 1024\}$  QAM constellation.  When the BER is $10^{-3}$, the $<E_{b(elec)}/{\xi^2}>$ and the $<E_{b(opt)}/{\xi^2}>$ against the normalized bit rate for  the different waveform are shown in Fig.\ref{P1_gauss} and Fig.\ref{P2_gauss}. It could be seen that the performance of the ACO-OFDM waveform and that of the U-OFDM waveform overlap each other. For the  $<E_{b(elec)}/{\sigma^2}>$ in Fig.\ref{P1_gauss}, the DCO-OFDM waveform with the best DC bias level has the best power efficiency.  The second and first type of  X-OFDM waveform outperforms the ACO-OFDM/U-OFDM waveform, when the normalized bit rate is larger than 1.13 and 1.73 bits/s/Hz. For the  $<E_{b(opt)}/{\sigma^2}>$ in Fig.\ref{P2_gauss}, the best waveform is the DCO-OFDM waveform with the best DC bias level. The second and first type of X-OFDM waveform outperforms the ACO-OFDM/U-OFDM waveform, when the normalized bit rate is larger than 1.82 and 2.32 bits/s/Hz.

Moreover, as shown in Fig.\ref{P1_gauss}, the required $<E_{b(elec)}/{\sigma^2}>$ for the two types of X-OFDM waveform with 4096-QAM are 36.2 and 37.2 dB to achieve the BER of $10^{-3}$. The normalized bit rate is the bits carried by used sub-carrier $\log_2 4096$ times the spectral efficiency $1/3$, whose value is 4 bits/s/Hz. Referring to the definition of $<E_{b(elec)}/{\sigma^2}>$ in equation (\ref{ref_snr}), the required $<E(X^2)/{\sigma^2}>$ for the two types of X-OFDM waveform with 4096-QAM are calculated as 42.2 and 43.2 dB. The simulations in Fig.\ref{P1_dep_40} and Fig.\ref{P2_dep_40} have $<E(X^2)/{\sigma^2}> = 40 \text{dB}$, which leads to that the 4096-QAM could not achieve the BER of $10^{-3}$. We note that $<E(X^2)/{\sigma^2}>$ is used for the fair comparison with the same channel condition, and $<E_{b(elec)}/{\xi^2}>$, $<E_{b(opt)}/{\xi^2}>$, $<E_{b(elec)}/{\sigma^2}>$ and $<E_{b(opt)}/{\sigma^2}>$ are utilized for the fair comparison of the power efficiency with the same data rate.


\section{Conclusion}\label{section5}
In this paper, two types of X-OFDM waveform  are proposed for the OWC  to satisfy the real and non-negative constrains. The related system architecture, signal processing, spectral efficiency, computational complexity, motivation and comparison, numerical simulations considering the OWC channel are comprehensively studied. Without the DC bias, the proposed two types of X-OFDM waveform achieve 1/3 spectral efficiency and lower PAPR than the ACO-OFDM/U-OFDM waveform. The first type of X-OFDM waveform uses non-successive demodulation scheme with computational complexity as $\mathcal{O}\left( 3N \text{log}_2(N) \right)$. The second type of X-OFDM waveform uses successive demodulation scheme with computational complexity as $\mathcal{O}\left( 5N \text{log}_2(N) \right)$ and higher power efficiency. With only the signal-dependent noise, the second type of X-OFDM waveform has the best electrical and optical power efficiency, when the normalized bit rate is larger than 0.9 and in [1.88, 3.46] bits/s/Hz, respectively. With only the signal-independent noise and without the DC bias level,  the second and first type of X-OFDM waveform outperforms the ACO-OFDM/U-OFDM waveform, when the normalized bit rate is larger than 1.13 and 1.73 bits/s/Hz for the electrical power efficiency, and 1.82 and 2.32 bits/s/Hz for the optical power efficiency. The proposed two types of X-OFDM waveform could be standardized and utilized in the OWC system in the branch without the DC bias.

\begin{IEEEbiography}[{\includegraphics[width=1in,height=1.25in,clip,keepaspectratio]{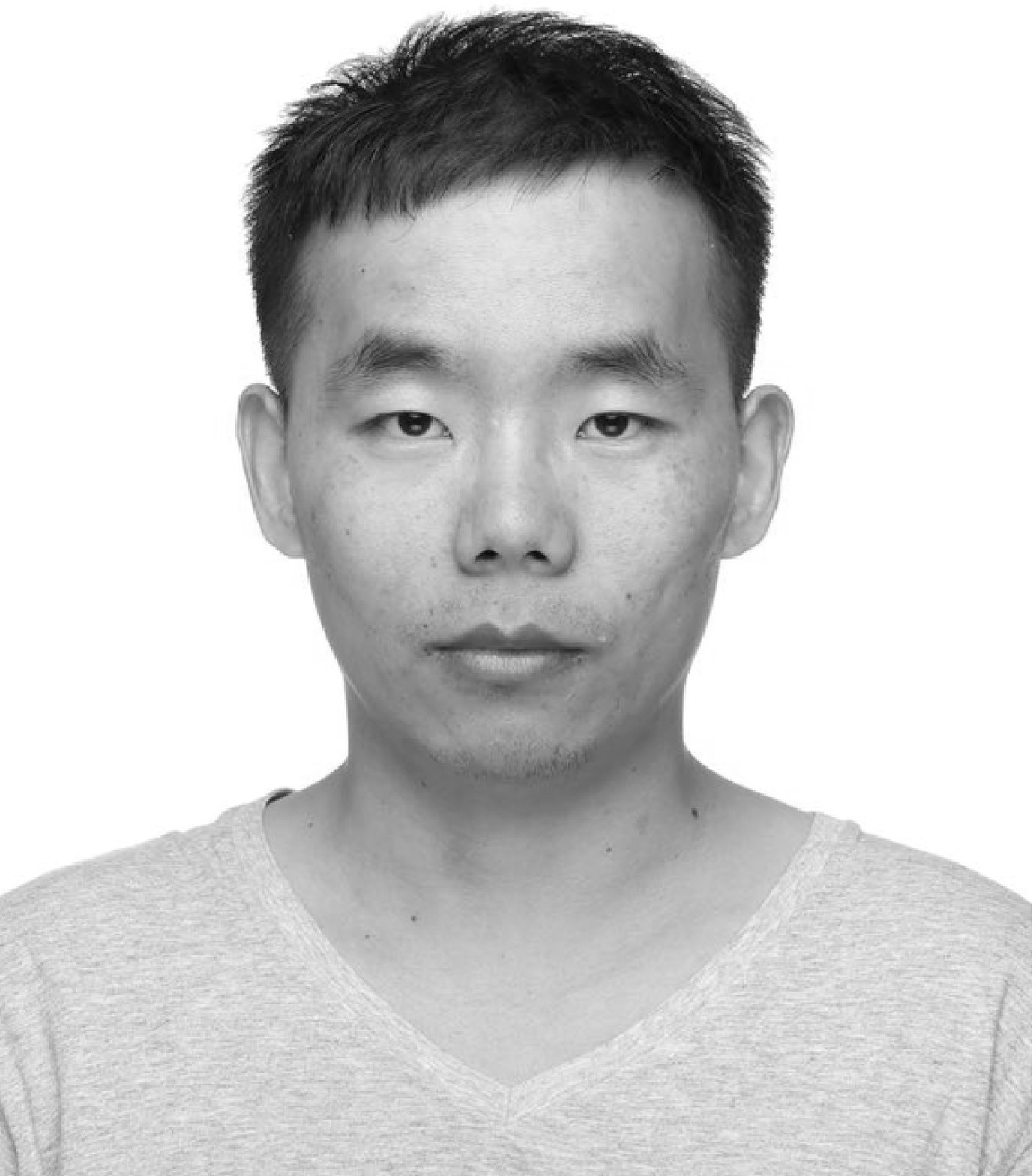}}]{Xu Li}
received the B.S. and Ph.D. degrees in electrical and electronics engineering
from the University of Science and Technology of China (USTC),
in 2010 and 2015, respectively. From 2013 to 2014, he was a Visiting Ph.D.
student at the Department of Electrical Engineering and Computer Science,
Northwestern University, U.S. His research involves various architectures of
radio access network in 5G, especially the end to end network slicing. He has
a rich wireless research experience including ultra-wide band (UWB) chips,
interference alignment, relay networks, stochastic geometry, and public safety
wireless broadband networks. His current research topics include microwave
wireless communication, radio over fiber communication, and optical wireless
communication.
\end{IEEEbiography}

\begin{IEEEbiography}[{\includegraphics[width=1in,height=1.25in,clip,keepaspectratio]{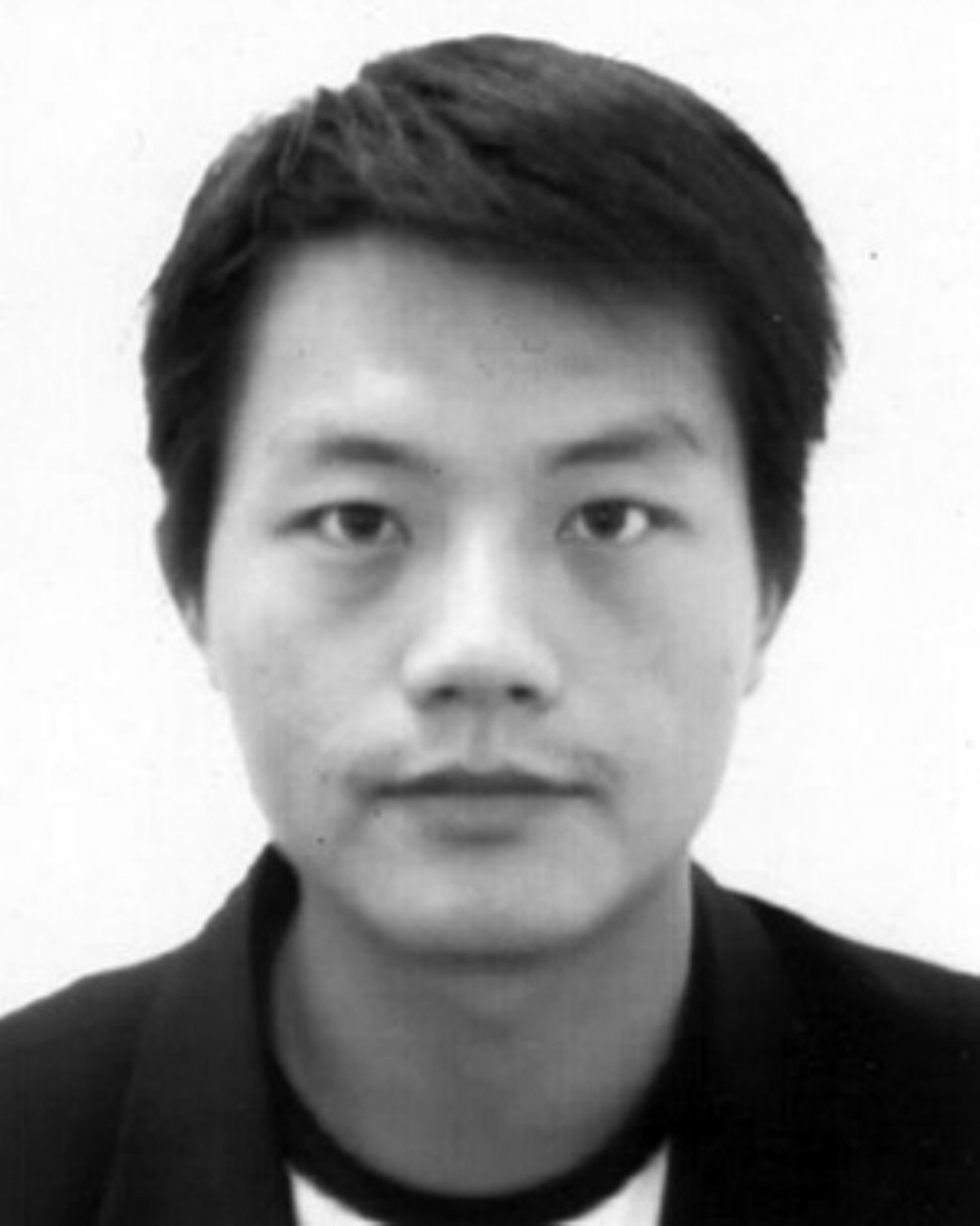}}]{Jingjing Huang}
received the B.Eng. and M.S. degrees in electronic engineering from Tsinghua University, Beijing, China, in 2003 and 2006, respectively. After that he was admitted by University College London, UK and received his Ph.D. degree in electronic engineering in 2010. After graduation he joined Novella Satcomms in Leeds, UK and worked as a RF engineer for 2 years. In 2013 he joined Passive-Eye Ltd and developed a self-powering GPS tracking system. In 2015 he joined TP-Link technologies Co. LTD in Shenzhen, China and worked as a system engineer for one year. He is now with Huawei Technologies Co. Ltd and working at the Central Research Institute, 2012 Labs as a research engineer. His research interests include antennas, array signal processing, RF/microwave circuits, energy harvesting, low power wireless transceiver, backscatter communication, satellite communication and optical wireless communications.
\end{IEEEbiography}

\begin{IEEEbiography}[{\includegraphics[width=1in,height=1.25in,clip,keepaspectratio]{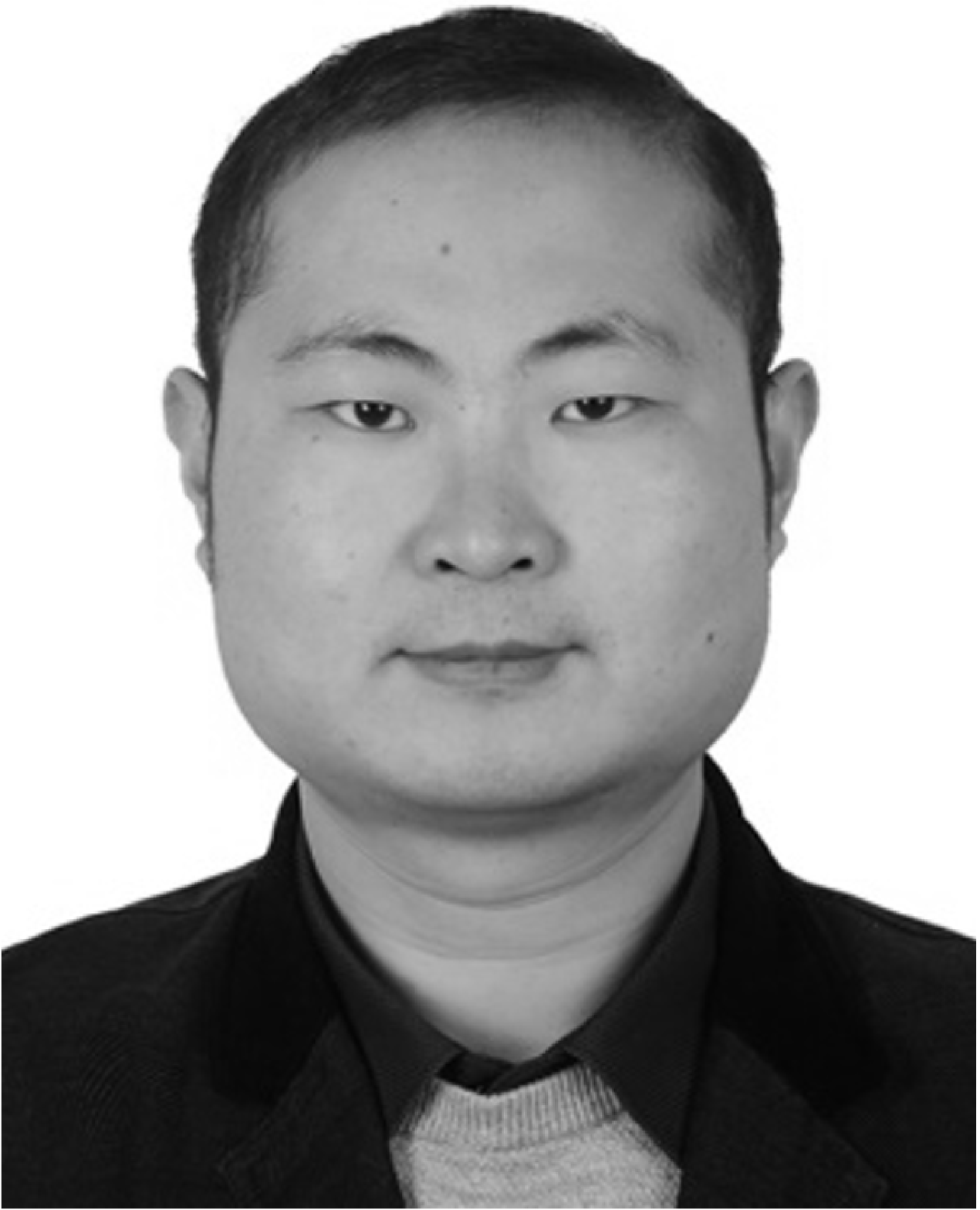}}]{Yibo Lyu}
received the M.Sc. degree in information and signal processing from the Chongqing University of Posts and Telecommunications in 2011 and the Ph.D. degree in circuits and systems from Xiamen University in 2016. His research interests include channel coding, joint source and channel coding, chaotic communications and nonlinear compensation. His current research projects including the modulation and coding methods for wireless optical communication system and the signal processing algorithms for radio over fiber system.
\end{IEEEbiography}

\begin{IEEEbiography}[{\includegraphics[width=1in,height=1.25in,clip,keepaspectratio]{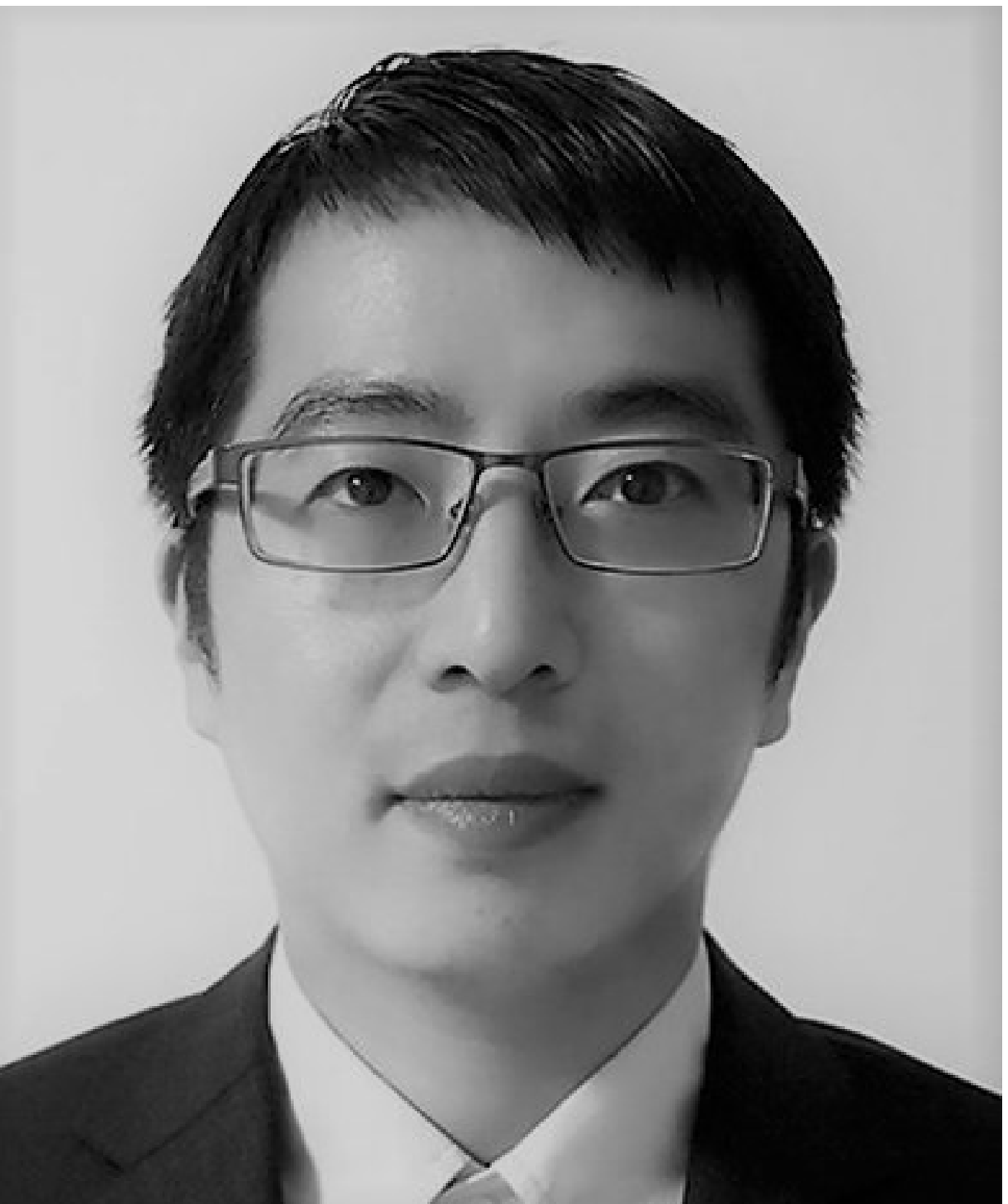}}]{Rui Ni}
 received the B.Sc. degree and Ph.D. degree in electrical and electronics engineering from University of Science and Technology of China, in 2006 and 2011, respcetively. From 2011 to now, he was a principal engineer in the Central Research Institute, 2012 Laboratory of Huawei Tech. Co. Ltd. His research involves various architectures of radio access network and core network, such as LTE, UMTS, GSM, WiMAX and Wi-Fi. He has a rich communication network related experience in engineering practice. His current research interests include Beyond 5G, electromagnetic information theory, and new waveforms.
\end{IEEEbiography}

\begin{IEEEbiography}[{\includegraphics[width=1in,height=1.25in,clip,keepaspectratio]{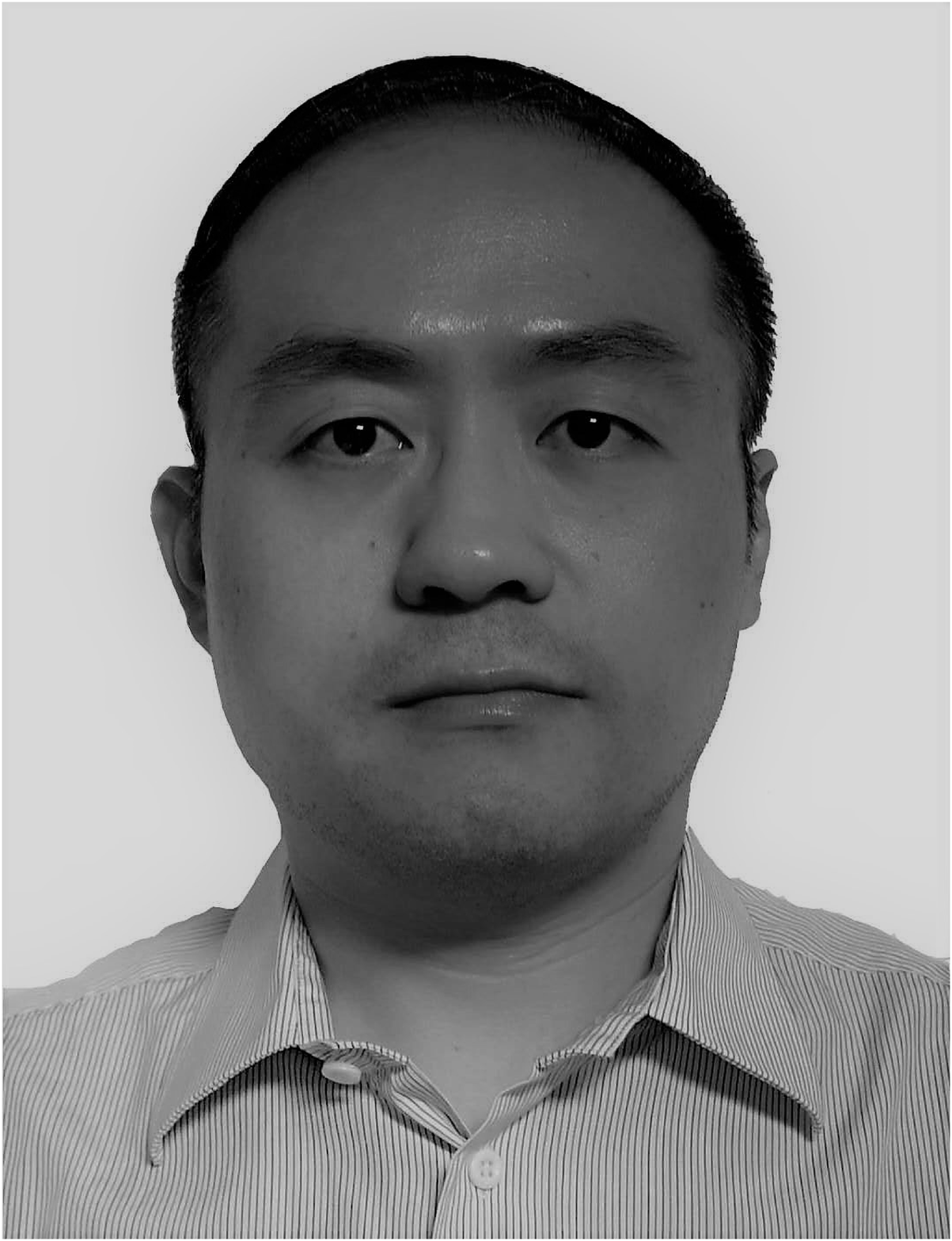}}]{Jiajin Luo}
received the B.S. and M.S degrees in electronic and communication engineering from Harbin Institute of Technology in 1999 and 2001, respectively. His research involves GNSS and indoor localization, satellite communication, cellular communication . He is experienced in system and signal design of GNSS, high dynamics signal acquisition and tracking, ultra-high sensibility receiver. His current research topics include positioning, integrated sensing and communication, joint communication and navigation.
\end{IEEEbiography}

\begin{IEEEbiography}[{\includegraphics[width=1in,height=1.25in,clip,keepaspectratio]{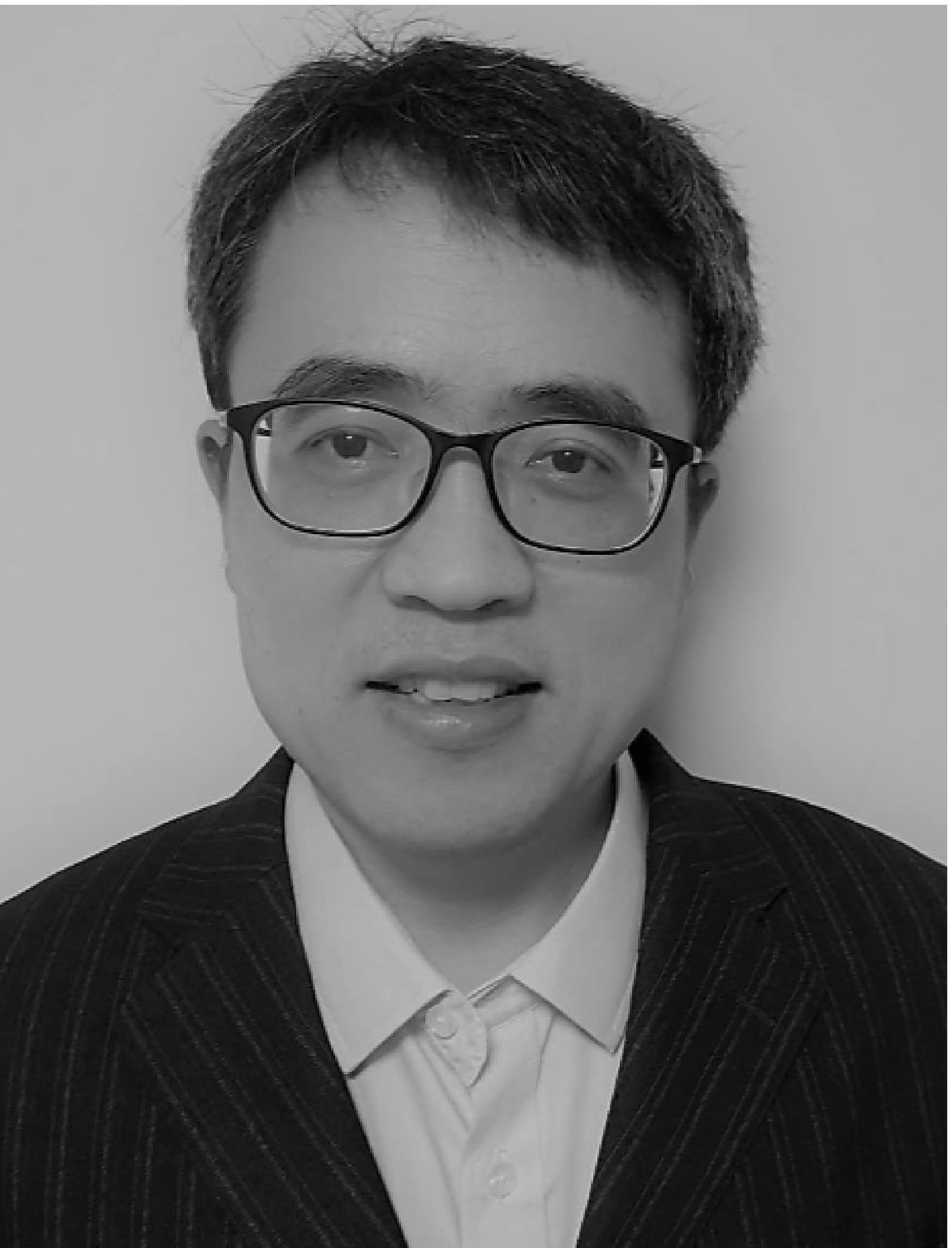}}]{Junping Zhang}
is a principal engineer of Huawei Wireless technology lab, Huawei, Shenzhen China. He is mainly engaged in optical wireless technology research and focuses on the application of optical spectrum in next-generation wireless communications. His research scope covers new optical sources, detectors, and new air interface technologies. He has more than 15 years of experience in wireless communications technology research and more than 20 granted patents.
\end{IEEEbiography}




\end{document}